\documentclass[a4paper,12pt]{article}
\usepackage[utf8]{inputenc}
\usepackage[english]{babel}
\usepackage{amsmath, amssymb,graphicx}
\usepackage{yfonts}
\usepackage{caption}
\usepackage[normalem]{ulem}
\usepackage{subcaption}
\pdfoutput=1 
\usepackage{jheppub} 
\newcommand{\be}{\begin{equation}}
\newcommand{\ee}{\end{equation}}
\newcommand{\bea}{\begin{eqnarray}}
\newcommand{\eea}{\end{eqnarray}}
\newcommand{\nn}{\nonumber}

\title{Holography for Heavy Ions Collisions at LHC and NICA }

\author[a]{Irina Ya. Aref'eva,}

\affiliation[a]{Steklov Mathematical Institute, Russian Academy of Sciences,\\Gubkina str. 8, 119991, Moscow, Russia}

\emailAdd{arefeva@mi.ras.ru}

\abstract{ This is a contribution for the Proceedings of 5th International Conference on New Frontiers in Physics (ICNFP 2016), held at Crete, 6-14 July 2016.
Our goal is  to obtain phenomenologically reliable insights for the physics of the  quark-gluon plasma (QGP)
from the holography.
I briefly review how in the holographic setup one can describe the QGP formation in heavy ion collisions 
and how to get quantitatively the main characteristics of  the QGP formation -- the total multiplicity and the thermalization time.
To fit the experimental form of dependence of total multiplicity on energy, obtained at LHC,  we have to deal with a special anisotropic holographic model, related with the Lifshitz-type background.

Our conjecture is that this Lifshitz-type background with non-zero chemical potential can be used to describe future data expected from NICA.
In particular, we present the results of calculations of the holographic  confinement/deconfinement  phase transition in the 
$(\mu,T)$ (chemical potential, temperature) plane  in this anisotropic background and show the dependence of the transition line on the orientation of the quark pair. This dependence leads to a non-sharp character of physical confinement/deconfinement phase in the $(\mu,T)$-plane. We use the bottom-up soft wall approach incorporating 
quark confinement deforming factor  and  vector field providing the non-zero chemical potential. In this model we also estimate the holographic photon
production.
}


\begin{document}
\maketitle
\flushbottom
\newpage

\section{Introduction}
\label{intro}
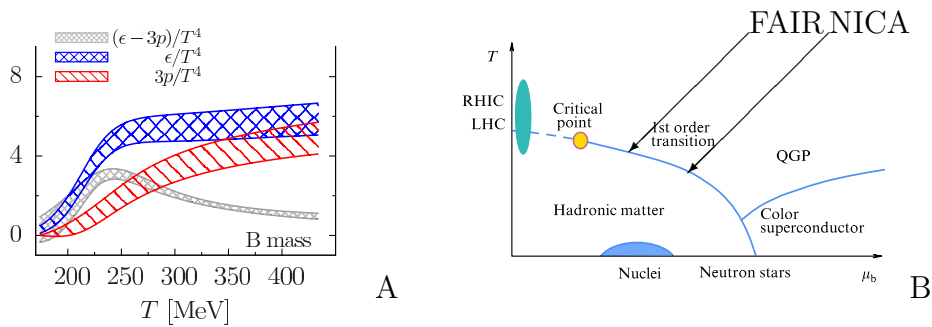
\begin{figure}[h]
\centering
\centering \begin{picture}(185,140)
\put(-80,0){\includegraphics[width=4.5cm]{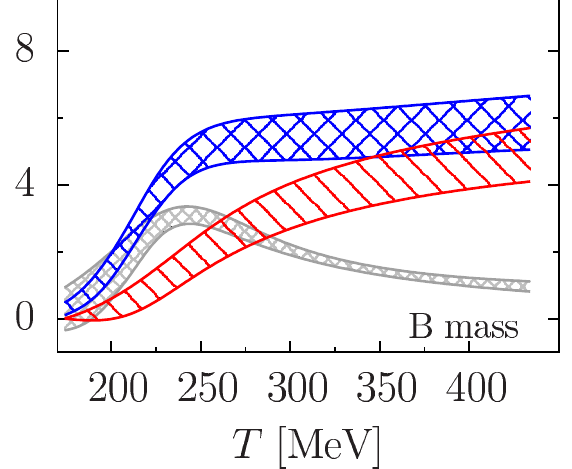}}
 \put(-60,90){\includegraphics[width=2cm]{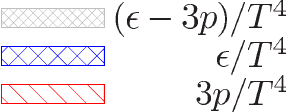}}
 \put(90,10){\includegraphics[width=6cm]{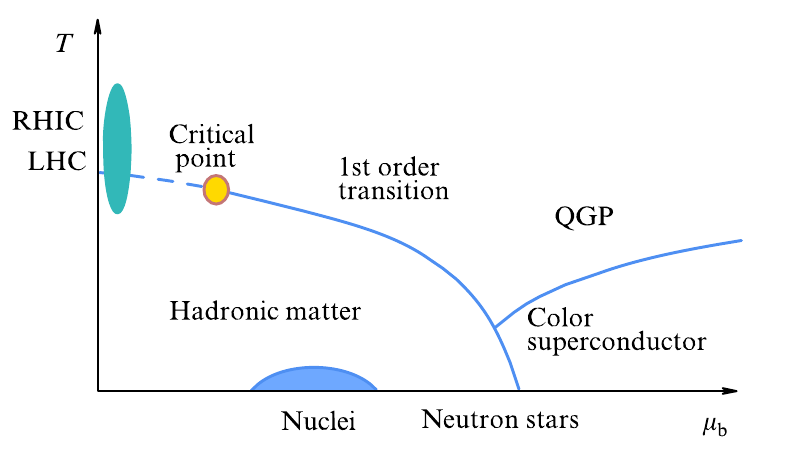}}  
  \put(200,110){\vector(-1,-1){45}}
   \put(230,110){NICA}
    \put(230,110){\vector(-1,-1){53}}
     \put(200,110){FAIR}
  \put(60,10){A}\put(260,10){B}
 \end{picture}
\caption{A. The trace anomaly, energy density and pressure for two flavors
of twisted mass Wilson fermions at $m_\pi=360$~MeV. The plot from \cite{Bazavov:2015qsa}. B.The QCD phase diagram}
     \label{Fig: TQCD} 
\end{figure}

Quark Gluon Plasma (QGP) produced in heavy ion collisions (HIC) at RHIC and LHC is a new form of matter
formed from quarks and gluons at high temperature \cite{RHIC,Aamodt:2010pb}.
It is  a strong coupling fluid \cite{fluid2},  or an "operator boiling fluid"
\cite{Feinberg}. This makes perturbative methods inapplicable to study properties of QGP and especially 
  it formation. The lattice QCD, the  powerful tool  of  non-perturbative study, is  formulated in the Euclidean spacetime and  cannot be applied to QGP formation, since this is 
 time depending phenomena (cf.\cite{Wilson}).
This gives a strong motivation  to study the process of QGP formation in HIC through the gauge/string duality.
The gauge/string duality (or AdS/CFT correspondence) is directly applicable only for a very special 4-dimensional theory, namely, $\mathcal{N}$=4 SUSY  Yang Mills theory \cite{Malda-rev}.  This  theory is conformally invariant on the quantum level and the conformal invariance is the main request for  the AdS/CFT application. Of course, the real  QCD is not conformally invariant, however,  as we know from lattice calculations, the high-temperature QCD is. More precise, the deviation from conformality  decreases  as the temperature increases, see Fig.\ref{Fig: TQCD}.A.   The agreement of the  shear viscosity to entropy density ratio of the QGP formed at RHIC and LHC with its holographic calculation \cite{PSS}  strongly
supports  the idea to use holography to study  physics of QGP formed in HIC \cite{}.

 In Fig.\ref{Fig: TQCD}.B the QCD phase diagram is presented. The phase diagram of QCD is not well known either experimentally or
theoretically.
A commonly conjectured form of the phase diagram,  temperature T vs quark chemical
potential $\mu$, is shown in Fig.\ref{Fig: TQCD}.B.
The  chemical potential $\mu$ is a measure of the imbalance between quarks and
antiquarks in the system. The phase transition is not sharp and it is supposed to be the 1-st order.

Ordinary  nuclear matter in this diagram is  at $\mu$ = 310 MeV and
T close to zero. If we increase the quark density, i.e. increase
$\mu$, keeping the temperature low, we go into a phase of 
more and more compressed nuclear such as matter 
neutron stars.
Above the (blue on the on-line version of the paper) smeared line
  there is a transition to the quark-gluon plasma.
  At ultra-high densities
   one expects to find the
 phase of color-superconducting quark
matter. 
In ultra-relativistic heavy ion collisions one studies this
matter  in the regime of extreme energy density. 
In Fig.\ref{Fig: TQCD}.B.  the
typical values of $\mu$ and $T$ in heavy-ion collisions (RHIC and LHC) are shown by a filled (cyan) region near the T-axis. The regions expected to be available at NICA (Nuclotron-based Ion Collider fAcility) \cite{NICA} and FAIR (Facility for Antiproton and Ion Research) are indicated by arrows.

 The  exact dual description of the real  QCD is unknown, but  holographic
 QCD models that fit  perturbative (two loops $\beta$-function) and lattice QCD  results (in particular, the quark confinement potential) have been proposed \cite{Gursoy:2008za,Gubser-rew}.
 Using these models several static properties of QGP  have been reproduced \cite{solana}.

 The description of the QGP formation in HIC is a difficult  subject, since it supposes to study a complicated real time 
 phenomena -- thermalization.  We also do not  know much from experiments about the details of the QGP formation in HIC, one can just estimates the time  of QGP formation  as well as the total multiplicity (there are arguments that the main part of particles is produced 
 during the QGP formation) \cite{1108.6027,1512.06104}.   The QGP formation  has been the subject of the  active studies within holographic approach in last years (see
 \cite{IA,DeWolf,Wilke} and refs therein).  
 Initially  this problem was considered in AdS background  \cite{Gubser,AlvarezGaume:2008fx,Lin:2009pn,Albacete:2009ji,ABG,ABJ,Kovchegov:2009du} and  the total multiplicity 
 within this approach was estimated as
 \be\label{multy-ads}
 {\cal M}_{AdS}\sim s^{0.33},
 \ee
For  the improved holographic background  the estimation was \cite{KT}
\be\label{multy-IHqcd}
 {\cal M}_{IHQCD}\sim s^{0.22}(1+\log \,{\mbox{corrections}}),
 \ee
  The  experimental multiplicity
 dependence on energy \cite{1108.6027,1512.06104}  is 
 \be\label{multy-exp}
 {\cal M}_{LHC}\sim s^{0.155(4)},
 \ee
  and it has been shown in  \cite{APP} that the dependence \eqref{multy-exp}  requires an unstable background.
  
    \begin{figure}[h!]\centering
 \includegraphics[width=6cm]{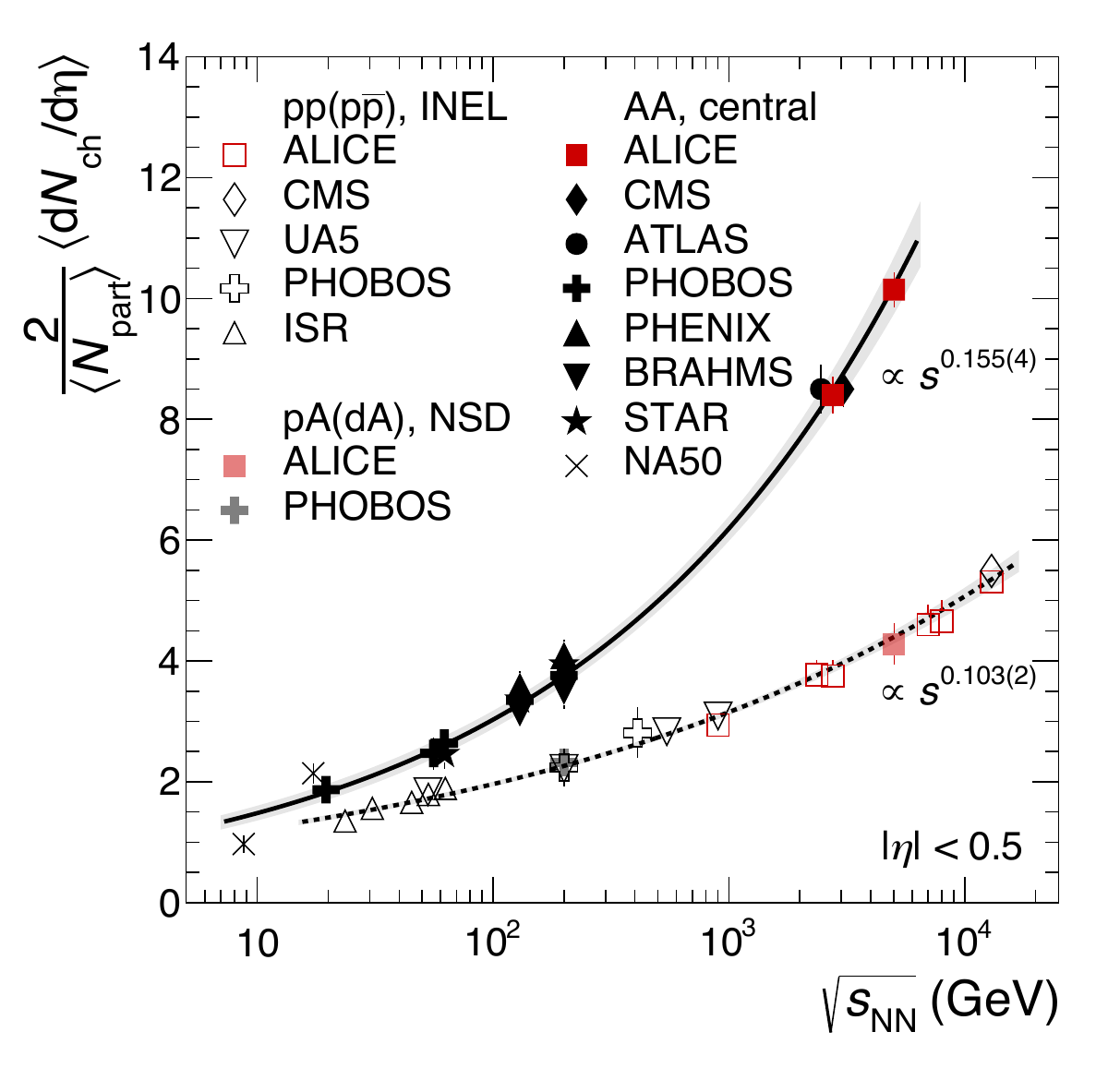}A$\,\,\,\,$$\,\,\,\,$$\,\,\,\,$$\,\,\,\,$$\,\,\,\,$$\,\,\,\,$$\,\,\,\,$$\,\,\,\,$$\,\,\,\,$ \includegraphics[width=3.5cm]{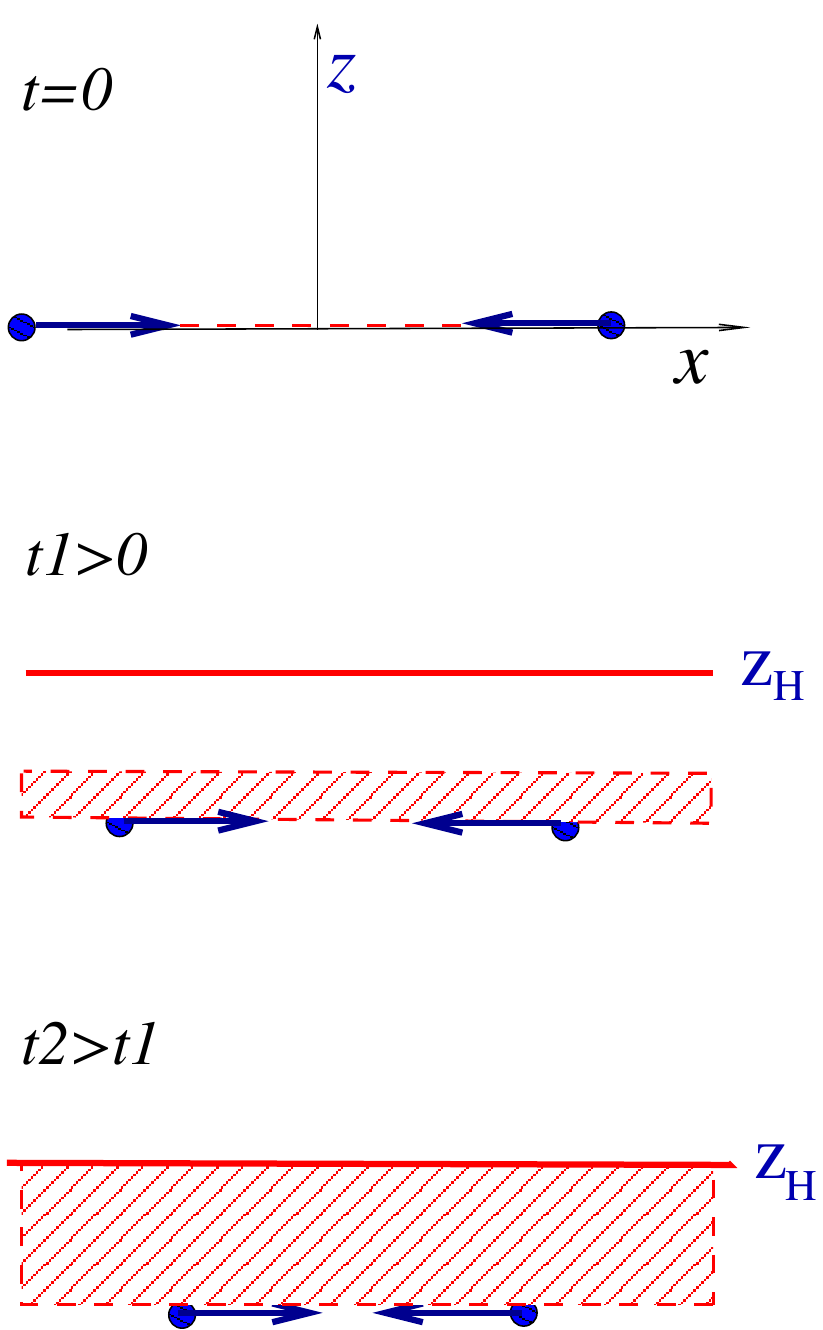}B
  \caption{A. The plot of the total number of charged particles versus
energy $\sqrt{s}$ for protons and $\sqrt{s_{NN}}$ per pair of nucleons for ions collisions. Plot from ALICE  collaboration \cite{1512.06104}. B. The schematic cartoon of the black hole formation in the bulk after collision of particles on the boundary.
}
        \label{Fig:Multy}
  \end{figure}
In \cite{Ageev:2014mma} it has been shown that  the  model that reproduces the Cornell potential
also gives a  correct energy dependence of multiplicities if we assume that the multiplicity is related with the
dual entropy produced  during a limited time period.
However in this consideration there is a limitation on the possible energy of colliding shock walls\cite{Ageev:2014mma}. Since in this consideration we have used a more or less general isotropic background
reproducing AdS at UV and confinement at IR, we can think  that the  assumption
about the isotropic background prevents to reproduce \eqref{multy-exp} at  high energy.  

In \cite{AG} we have considered a special {\it anisotropic} backgrounds, parametrized by a 
 paranmeter $\nu$ and have got
\be\label{multy-exp-m}
 {\cal M}_{\nu}\sim s^{1/(\nu+2)},
 \ee
 and therefore to get an estimation \eqref{multy-exp}  we take $\nu=4.45
 $. Note, that $\nu=4$ gives 
 $
 {\cal M}_{\nu=4}\sim s^{0.167}$,
 that is rather closed to \eqref{multy-exp}. $\nu=1$ corresponds to the AdS case.

Let us note, that the idea to use anisotropic background in the holographic approach to QGP is not new. 
There were several proposals to deal with anisotropic backgrounds, see for example \cite{DiGi} and refs therein.
Up to couple years ego,  it has been believed  that the matter was in the pre-equilibrium period up to 1 fm/c and then  the QGP appears and it is
isotropic.  However   now it  is  believed  that    the QGP  is created after a very short time after collision,
$\tau _{therm}\sim 0.1 fm/c$,
and it is anisotropic ("anisotropic" means  a spatial anisotropy)  for  a short time $\tau$ after collision,
$0<\tau_{therm}<\tau <\tau_{iso}$, and the time of local isotropization is about
$\tau_{iso}\sim 2 fm/c$ \cite{Strickland:2013uga}.

\section{Holographic Thermalization}
\label{sec-2}
\subsection{Thermalization}
Suppose we deal with a correlator of a time depending quantity ${\cal O}(t)$,  $<{\cal O}(t)>$. If  at large $t$
the system exhibits the behaviour
\be
<{\cal O}(t)>\underset{t\to \infty}\to<<{\cal O}>>,\ee
where
\be
<<{\cal O}>>=\mbox{tr}\left(e^{-\beta H} {\cal O}\right)\ee
one says that the system goes to the stationary state and this state is a thermal state, or in other words
 the system thermalizes. The simples thermalization process can be described by interaction with thermostat. For one oscillator  interacting with thermostat one can show the  thermalization explicitly.

\subsection{Thermalalization as a black hole creation in AdS}
\label{ssec-2.3}
According to the holographic scenario  the thermalization process in d-dimensional space-time 
can  be  understood  as horizon formation in a d+1 dimensional gravitational theory.
To initiate the process of BH formation one has to perturb the initial AdS metric.
According to AdS/CFT correspondence this deformation should be related with the deformation 
of the energy-momentum tensor of the boundary theory.
So the idea is to make some perturbation of AdS metric
that near the boundary mimics the heavy ions
collisions   and see what happens, see Fig.\ref{Fig:Multy}.B.

One can consider several deformations of AdS metric:
\begin{itemize} 
\item a deformation by adding colliding gravitational shock waves.

\item we can drop of a shell of matter with vanishing rest mass ("null dust"),

\item we can study the toy 3-dim model with colliding ultra relativistic particles
\end{itemize}

It seems natural to think about an ultrarelativistic nucleus as a shock
wave in 4-dimensional space-time. One can also assume that this shock has a profile function.
According to AdS/CFT correspondence,  this energy-momentum tensor
corresponds to  a shock wave metric in AdS$_5$ of the following form
\be ds^2=L^2
\frac{-du\,dv+dx_{\perp}^2+\phi(x_{\perp},z,L)\delta(u)\,du^2+dz^2}{z^2},
\label{shock}
\ee
here $u,v$ are light-cone variables, $u=x_0+x_{||}$,  $v=x_0-x_{||}$, $\phi(x_{\perp},z,L)$ is the profile function, $L$ is the scale parameter. For $\phi(x_{\perp},z,L)=0$ the metric \eqref{shock}
is nothing but the Poincare metric for AdS$_5$.
For two colliding nuclei corresponds the following metric in AdS$_5$
\be
ds^2=L^2\frac{-du\,dv{+}dx_{\perp}^2{+}
\phi_1(x_{\perp},z,L)\delta(u)\,du^2{+}
\phi_2(x_{\perp},z,L)\delta(v)\,dv^2{+}dz^2}{z^2}.
\label{two-shock}\ee
 It is interesting to note that gaussian regularization of the point shock wave profile function $\phi(x_{\perp},0)$ exactly corresponds to the Woods-Saxon profile
for the nuclear density. This correspondence  gives us L equal to 4.3 fermi for gold and
L equal to 4.4 fermi for lead.

It is natural to ask the question: can we guarantee the black hole formation under collisions of these shock waves.
If the answer is "yes", then we take the entropy of this formed black hole
as a multiplicity of particles production. This idea has been explored in several papers \cite{Gubser,AlvarezGaume:2008fx,Lin:2009pn,Albacete:2009ji,ABG,ABJ,Kovchegov:2009du,KT}.
The idea to relate multiplicities with the thermodynamic characteristic of the media produced as a result of a collision of particles comes back to Landau and Fermi \cite{Landau,Fermi,Pomeranchuk}.

 \begin{figure}[h!]\centering
  \begin{picture}(185,140)
\put(-80,0){\includegraphics[width=4.5cm]{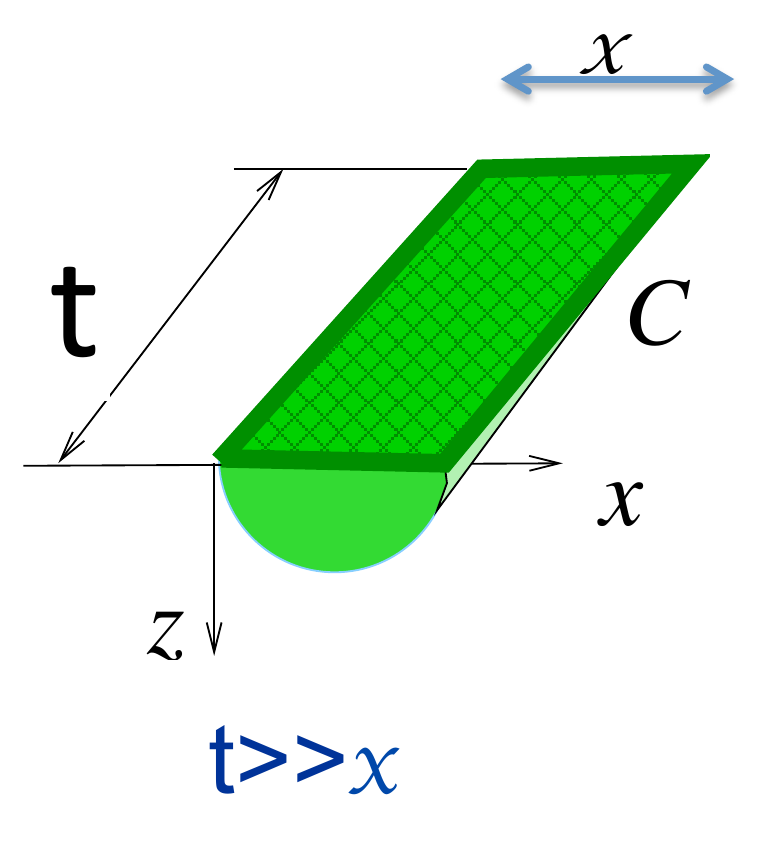}A}
 \put(80,20){\includegraphics[width=4.5cm]{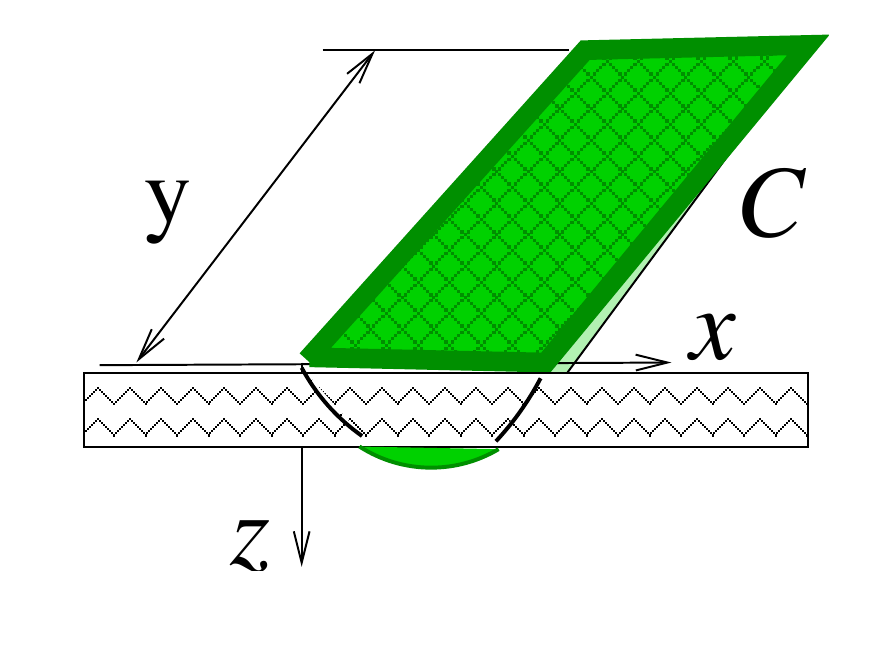}B}
 \end{picture}
  \caption{ A. Wilson loop in the static background. B. Wilson loop in the Vaidya background
}
        \label{Fig:WL}
  \end{figure}
\newpage
\subsection{Thermalization time}
Thermalization time depends on the physical quantities that are expected to thermalized.
This issue has been discuss in details in our talk \cite{Aref'eva:2016doe}  and is not discussed here, see also 
\cite{IYaA,AAGG} for estimations of thermalization time of different variables in the Lifshitz-type background.

\section{Physics in the Lifshitz-type background}
\subsection{Lifshitz-type background}
In \cite{AG} we have studied  collisions of shock waves in the anisotropic background
 \be\label{Lif-like}
ds^{2} =2\pi\alpha'\left(\frac{-dt^{2} + dx^{2}}{z^{2}} + \frac{dy^{2}_{1} + dy^{2}_{2}}{z^{2/\nu}} +  \frac{d z^{2}}{z^{2}}\right),
\ee
where $\nu$ is the critical exponent.  

After collision the black hole is produced and the stationary black hole metric in
this background is \cite{AGG}
\be\label{ds-AGG}
ds^{2} = z^{-2}\left(-f( z)dt^{2} + dx^{2}\right) + z^{-2/\nu}(dy^{2}_{1} + dy^{2}_{2})  + \frac{d z^{2}}{ z^2 f(z)},
\end{equation}
where the blackening function is
\be \label{f-AGG}
f = 1- m z^{2/\nu + 2}.
\ee

Motivated by \cite{9902170,1001.4414,1008.3116} we deform this background as
  \footnote{The problem of finding of shock waves in the background \eqref{ds-aniz-con-chem} and estimation of the trapped surface in this background is  not solved yet, compare with \cite{ABJ,ABP}}
\be\label{ds-aniz-con-chem}
ds^{2} = \frac{b(z)}{z^{2}}\left(\left(-f( z)dt^{2} + dx^{2}\right) + z^{2-2/\nu}(dy^{2}_{1} + dy^{2}_{2})  + \frac{d z^{2}}{ f(z)}\right),
\ee
where the blackening function takes into account a non-zero chemical potential
\be
 \label{f-q-zh-nu}
f = 1-  (\frac{1}{z_h^{2/\nu+ 2 }} +q^2 z_h^2)z^{2/\nu + 2}+q^2 z^{2/\nu + 4}
\ee
and $b(z)$ is a factor that provides the quark confinement in the isotropic case \cite{AZ}
\bea
\label{b-AZ}
b(z)&=&\exp(\frac{c z^2}{2})
\eea
and the zero-component of the vector field is given by
\be\label{A0}
 A_0(z)=i \left(\mu-\frac{ \sqrt{3}\, q}{2 c} \left(1-e^{-c z^2}\right) \right)\,.
\ee
According  the AdS/CFT dictionary, the boundary value $A_0(0)$ is related to  the quark chemical potential: 
\be
A_0(0)=i\mu
\ee
The vanishing of $A_0(z)$ at the horizon, $A_0(z_h)=0$, sets a relation between the chemical potential and the charge of the black hole:
\be\label{mu}
 \mu=\frac{\sqrt{3} \, q}{ 2 c} \left( 1-e^{-cz_h^2} \right) \ee

\begin{figure}[h!]
\centering
\includegraphics[scale=0.4]{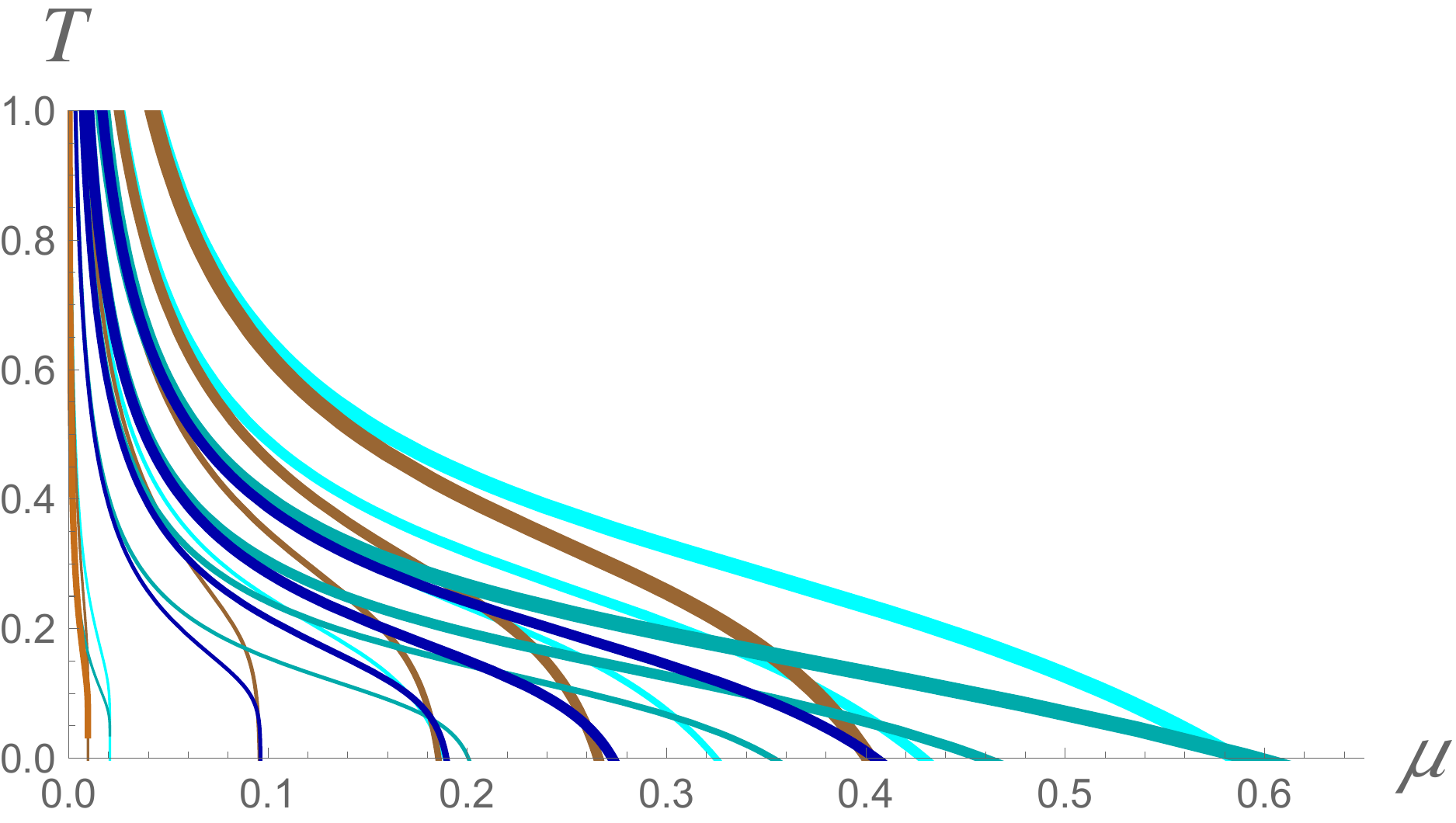}
\caption{
Relation between $T$ and   $\mu$ given by formula \eqref{temp} and \eqref{mu}  for isotropic (cyan  for $c=0.42$ and brown for $c=0.9$ lines)  and anisotropic case 
(darker cyan for $c=0.42$ and darker blue for $c=0.9$ lines ) with parameter $\nu=4$  and different values of the  charge $q$: $q=0.01,\,0.1,\,0.3,\,0.5$ (increasing  thickness of  the same color line corresponds to the increasing value of $q$). 
}\label{fig:phase-diagr-aniz}
\end{figure}

The temperature $T$ is defined by the relation
\be\label{temp}
 T=\frac{1}{4\pi}\left| \frac{df}{dz} \right|_{z=z_h}
 =\frac{1}{2\pi } \,\frac{\frac{1}{\nu} + 1}{z_h}\left(1 -
 \frac{Q_\nu^2}{2} \right),\ee
 where $Q_\nu=\sqrt{\frac{2}{\frac{1}{\nu} + 1}}q z_h^{\frac{1}{\nu} + 2}$. We can also present the dependence of $T$ on the chemical potential 
\be\label{T-zh}
T=\frac{\left(\frac{1}{\nu }+1\right)}{2 \pi  z_h} \left(1-\frac{c^2
   \mu^2 z_h^{\frac{2}{\nu }+4}}{3
   \left(\frac{1}{\nu }+1\right) \left(1-e^{-\frac{c
   z_h^2}{2}}\right)^2}\right)
\ee
 see Fig.\ref{fig:phase-diagr-aniz}.

\subsection{Temporal Wilson loop in  the  charged quark confinement background}
\subsubsection{Energy between quarks located along x-direction in  the  quark confinement background \eqref{ds-aniz-con-chem} }
The potential of the interquark interaction along x-direction can be extracted from the rectangular time-like Wilson loop of size $T\times X$, 
i.e. the loop in which one side is infinite along the time direction, and the other is along the $x$ direction
\be
 W(T,X)  = \langle {\mbox{Tr}}_F \,e^{i\oint_{T\times X} dx_\mu A_\mu} \rangle \sim
e^{-V(X) T },
\ee
$F$ means the fundamental representation. Following the holographic approach \cite{9803002,9803135,9803137} the expectation value of the Wilson loop in the fundamental representation calculated on the gravity side reads as:
\be\label{WC}
W[C]= e^{- S_{(xt)}},
\ee
 where $C=T\times  X$ in a contour on the boundary,   $S_{xt}$ is the minimal Nambu-Goto action of the string  hanging from 
 the contour $C$ in the bulk
\be\label{S-string}
S_{xt} = \frac{1}{2\pi \alpha'}\int d\sigma^{1} d\sigma^{2}\sqrt{-\det (h_{\alpha \beta})},
\ee
where $h_{\alpha\beta}$ is  the induced metric of the world-sheet
$
h_{\alpha\beta} = g_{MN} \partial_{\alpha}X^{M}\partial_{\beta}X^{N}$, $\alpha, \beta =0,1$,
 $g_{MN}$ is the background metric, $M, N =0,\ldots,4$,
 $X^{M}=X^{M}(\sigma^0, \sigma^1)$ specify the  string worldsheet with  the worldsheet coordinates $\sigma^{1}$, $\sigma^{2}$.  We parametrize the worldsheet as $X^0\equiv t=\sigma^0$ and $X^1\equiv x=\sigma ^1$ and 
 consider the static  configuration $X^4\equiv z=z(x)$. The action $S_{(xt)}$ is:
\be\label{S-tx}
S_{xt}
=\frac{T}{2\pi\alpha^{\prime}}\int \frac{b(z)} {z^2} \sqrt{f(z)+z^{\prime 2}}\,dx.
\ee
where $\prime$ means $\frac{d}{dx}$.
If we take $z'=0$ in \eqref{S-tx} we get the "potential"
\be\label{Vz}
V_x(z)=\frac{b(z)} {z^2} \sqrt{f(z)}\ee
Note that the form of the action \eqref{S-tx} is the same as for the isotropic case, and  information about anisotropy is stored only in the form of 
  blackening function $f$ \eqref{f-q-zh-nu}.
We consider the symmetrical parameterization $z(\pm \ell)=0$ and $z(0)=z_*$,  $z^\prime(0)=0$. The distance between  the two endpoints of the string  can be represented as
\be
\label{L1bmmm}
L_{x} =2\int^{z_{*}}_{\infty} \frac {dz}{z^{'}}=
2\int^{z_*}_{0}\frac{dz}{\sqrt{f(z)}\,\sqrt{\left(\frac{V^2_x(z)}{V^2_x(z_*)}-1\right)}}
\ee
We consider the case when $z_*<z_{h_{i_0}}$, where 
$z_{h_{i_0}}$ is the smallest  of the horizons, i.e. if $f(z_{h_i})=0$, then $z_{h_{i_0}}<z_{h_i}$. Therefore $f(z)>0$ for $0<z<z_*$. 

\begin{figure}[h!]
\centering
 \begin{picture}(185,130)
\put(0,0){\includegraphics[scale=0.35]{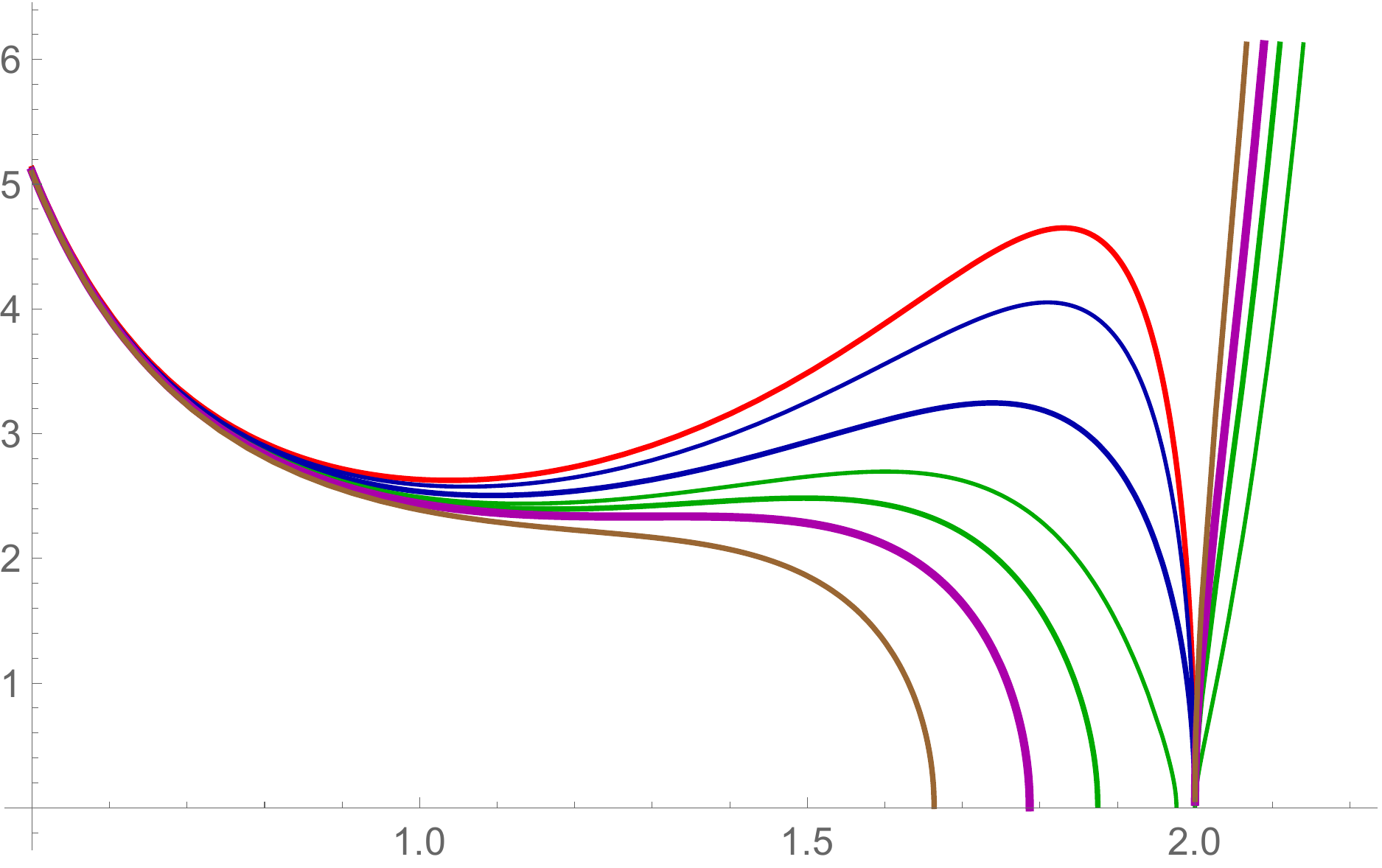}}
 \put(0,125){$V$}
  \put(180,0){$z$}
\end{picture}
 \caption{
 The potential $V(z)=V(z, c, q,\nu,z_h)$ given by  \eqref{Vz} with $b(z)$
 given by \eqref{b-AZ}  with $c=2$ and $f(z)=f(z,q,\nu,z_h)$ given by \eqref{f-q-zh-nu} with $\nu=1$  and $z_h=2$. The red line 
 shows $V(z)$ for  $q=0$. We see that the corresponding potential  has minimum and maximum.
 The blue lines 
 show $V(z)$ for    $q=0.1$ and  $q=0.15$. The corresponding potential have 
 minimum and maximum.   The green lines 
 show $V(z)$ for    $q=0.18$ and  $q=0.195$. The corresponding potential have 
 minimum and maximum and the second zero of the corresponding function $f(z)=f(z,q,1,z_h)$
 are on the left of $z_h=2$. The magenta line 
 show $V(z)$ for    $q=0.21$. The corresponding potential   has no
 local extremum  and  the second zero of  $f(z)$
 is at $z_{h_2}=1.784$.  For $q=0.21$ (the brown line)  the corresponding potential  has
no minimum and maximum on $0<z<z_{h_2}$ and the potential is the decreasing function at $z<z_{h_2}=1.662$}
\label{fig:pot-V-z-q-c2}
\end{figure}

\begin{figure}[h!]
\centering \begin{picture}(185,125)
\put(0,0){\includegraphics[scale=0.35]{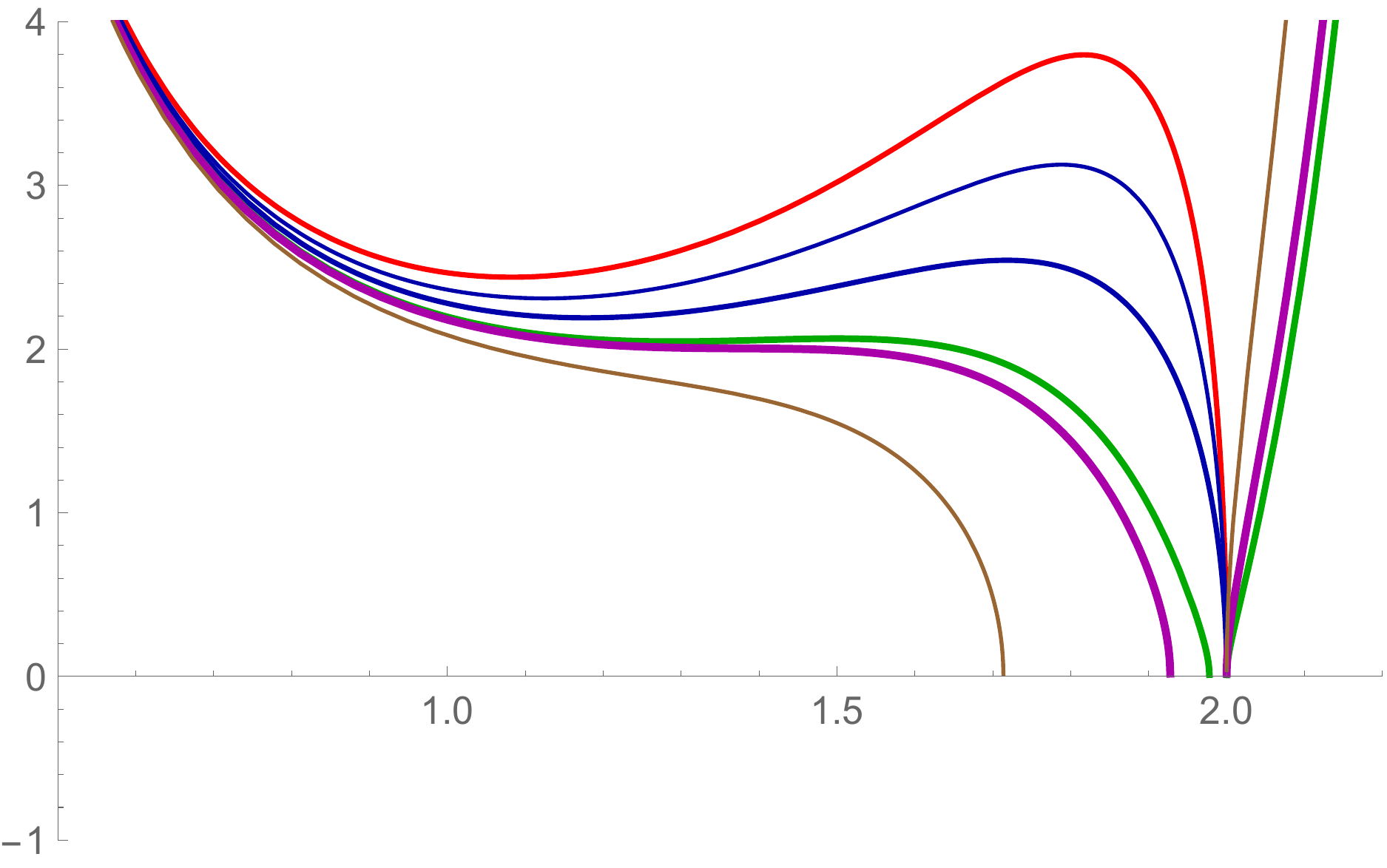}}
 \put(0,130){$V_x$}
  \put(180,20){$z$}
\end{picture}
 \caption{
 The potential $V_x(z)=V(z, c, q,\nu,z_h)$ given by  \eqref{Vz} with $b(z)$
 given by \eqref{b-AZ}  with $c=2$, and $f(z)=f(z,q,\nu,z_h)$ given by \eqref{f-q-zh-nu} with $\nu=4$  and $z_h=2$. The red line 
 shows $V(z)$ for  $q=0$. We see that the corresponding potential  has minimum and maximum.
 The blue lines 
 show $V(z)$ for    $q=0.15,0.2$. The corresponding potential have 
 minimum and maximum.  The green line 
 shows $V(z)$ for    $q=0.238$. The second zeros of the corresponding function $f(z)$
at $z=1.97$. The magenta line 
 shows $V(z)$ for    $q=0.245$ . The corresponding potential   has no
 minimum and maximum and  the second zero of the corresponding function $f(z)=f(z,q,1,z_h)$
 is at $z_{h_2}=1.893$. The brown line
 shows $V(z)$ for  $q=0.28$  and the corresponding potential has no
 minimum and maximum  on $0<z<z_{h_2}$, where $z_{h_2}=1.71$ is the second zero of the corresponding functions $f(z)$ .    The critical charge $q_{cr}|_{c=2,z_h=2}=0.245$
  }\label{fig:pot-V-z-q-c2-nu4}
\end{figure}

To get the energy of the string
 we subtract  the mass of the two free quark \cite{9803135,9803137} 
\bea
 \label{Ex}
\pi \alpha^{'}E_{x} &= &\int^{z_{*}}_{0} \frac{dz }{z^2}\left[\frac{b(z)V_x(z)}{\sqrt{V_x^2(z) -V_x^2(z_*)}}
-1
\right]
-\frac{1}{z_*}
+m^{\frac{\nu }{2 \nu +2}} e^{c\, m^{-\frac{2 \nu }{2 \nu
   +2}}}-\sqrt{\pi c} \,\text{erfi}\left(\sqrt{c}\,
   m^{-\frac{\nu }{2 \nu +2}}\right).
\label{WLxb-mm}
\eea
here $m$ is related with $z_h=\left(\frac{1}{m}\right)^{\frac{\nu }{2 \nu +2}}$.
To guaranty that the expression under the second square root in \eqref{L1bmmm} is positivity we have to integrate along 
the curve where the potential  $V_x(z)$ is a decreasing function. Depending on the parameters $c,z_h,\nu$
and  $q$  the function  $V_x(z)$ can have two extremal points, one minimum and one maximum, or one extremal point,  or even  have no extremal point at all
at the interval $0<z<z_{h_{i_0}}$, see Fig. \ref{fig:pot-V-z-q-c2} and Fig. \ref{fig:pot-V-z-q-c2-nu4}.

Therefore, as for isotropic cases \cite{Andreev:2006nw,Mia:2010zu,1008.3116,1201.0820,1506.05930,Ewerz:2016zsx,Fang:2015ytf} \footnote{Note, that the isotropic case with the confinement factor has been studied at zero chemical potential in \cite{Andreev:2006nw,Mia:2010zu} and with nonzero chemical potential in \cite{1008.3116,1201.0820,1506.05930,Ewerz:2016zsx,Fang:2015ytf}. }, we have two cases:
\begin{itemize} \item There is no extremal point in the interval $0<z<z_{h_{i_0}}$. This case corresponds to the deconfinement phase and dependence of $\ell$ on $z_*$ has the form as presented in Fig.\ref{fig:L-zs-above}.a).
We see that there are two branches indicated by blue and red color, the same $\ell$ can be obtained by two different $z_*$, except  $z_*=z_{*0}$ where $\ell$ reaches its  maximal value  $\ell_0$, i.e. $\ell=\ell(z_*)$ monotonically increases  from $\ell(0)=0$ to $\ell(z_{*_0})=\ell_0$ (the first branch shown by the blue color) and $\ell=\ell(z_*)$ monotonically decreases  from $\ell(z_{*_0})=\ell_0$ to $\ell(z_{h_0})=0$ (the second branch). In  Fig.\ref{fig:L-zs-above}.a) $z_{h_0}=z_h$. The plot in 
Fig.\ref{fig:L-zs-above}.b) shows the values of the energy between quarks located along x-direction as function of $z_*$ for two branches.
The corresponding values of the energy as functions $\ell$ are  presented in Fig.\ref{fig:L-zs-above}.c).  As shown in this plot,
the values of energy corresponding to the second branch  are larger than those for the first one.

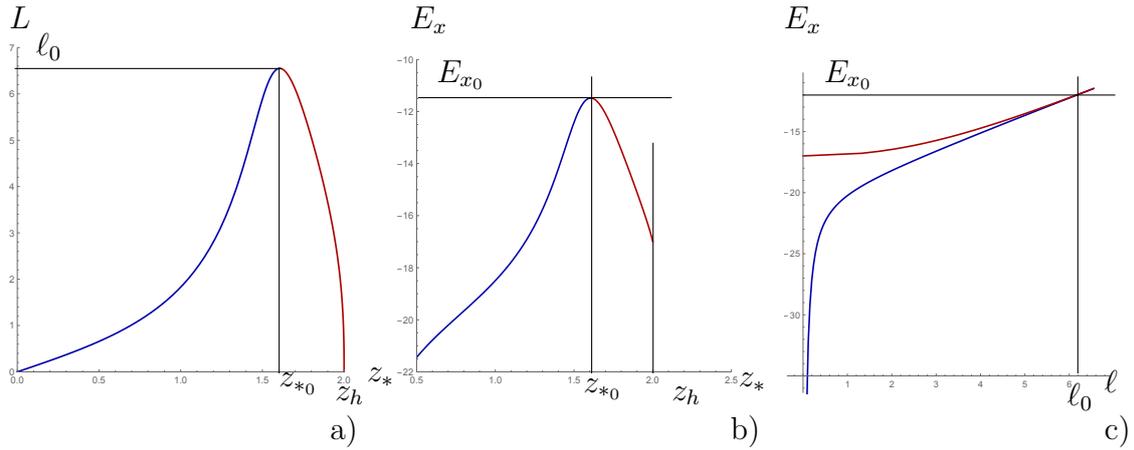
\begin{figure}[h!]
\centering \begin{picture}(185,150)
\put(-120,0){\includegraphics[width=4.5cm]{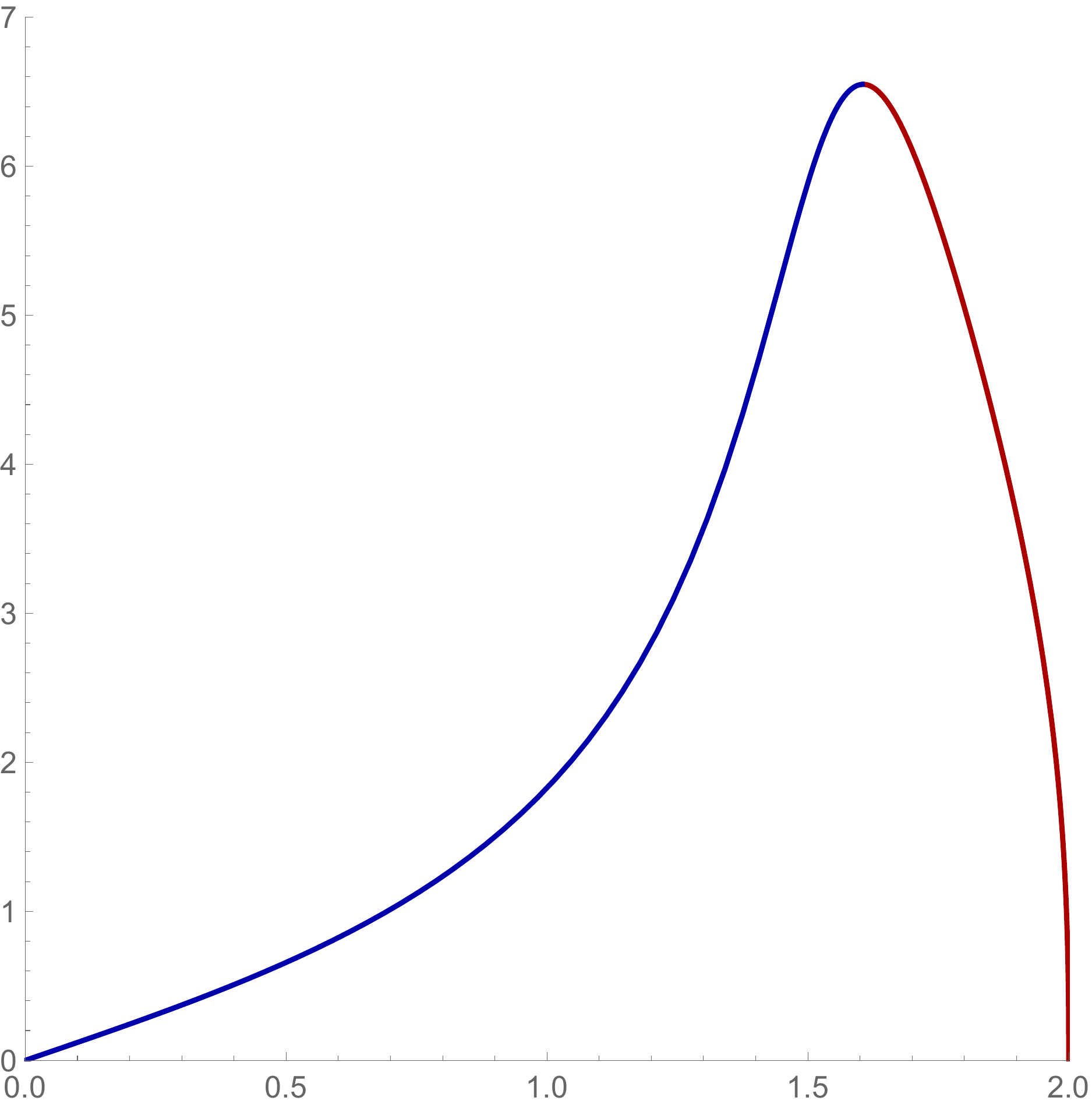}}
 \put(-120,135){$L$}
 \put(-19,5){\line(0,1){115}}
 \put(10,2){ $z_*$}
 \put(-19,0){$z_{*_0}$}
  \put(2,-5){$z_{h}$}
   \put(-118,120){\line(1,0){100}}
     \put(-110,125){$\ell _0$}
\put(25,0){\includegraphics[width=4.5cm]{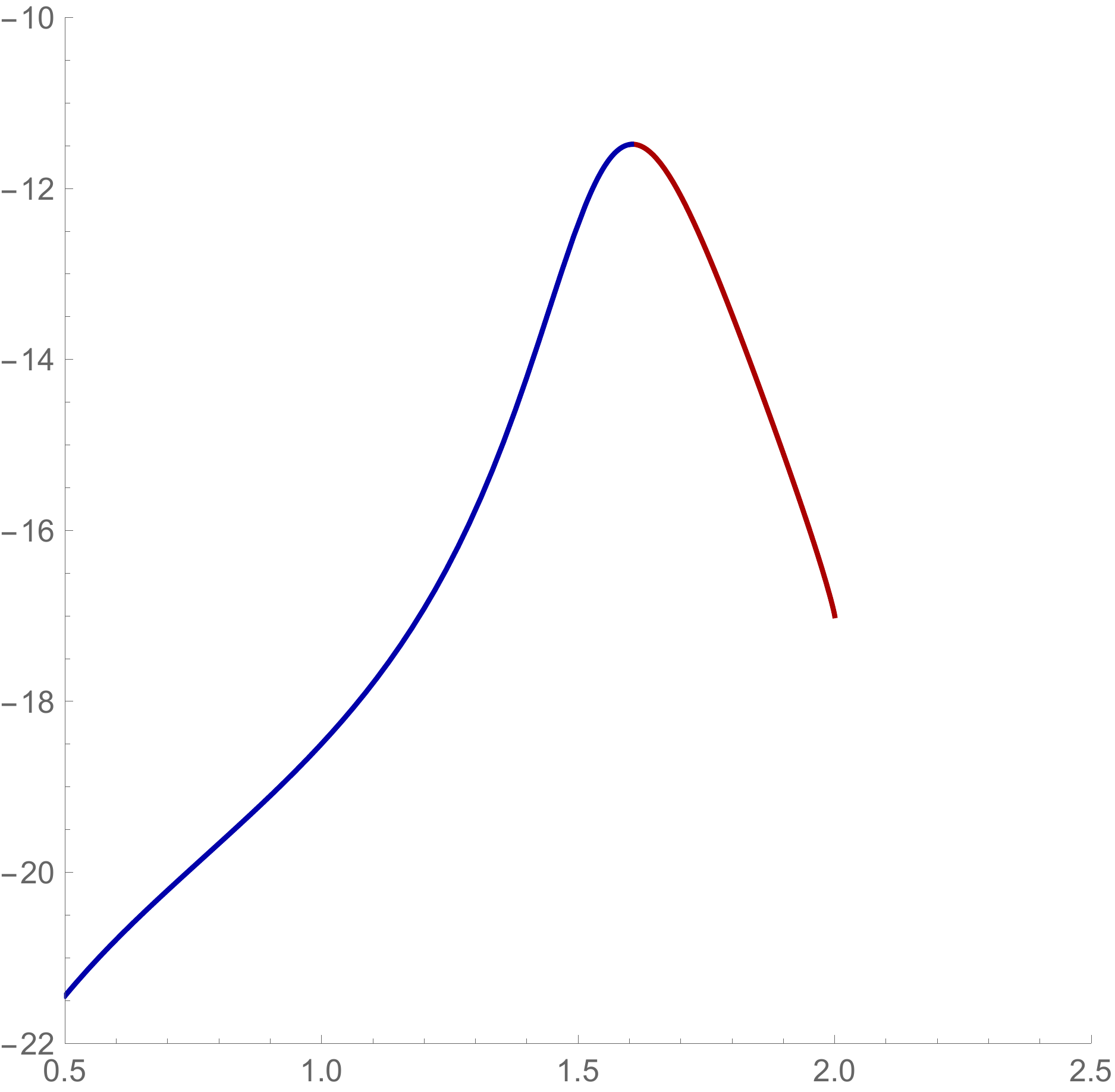}$z_*$}
 \put(30,135){$E_{x} $}
  \put(40,115){$E_{x_0} $}
  \put(98,5){\line(0,1){112}}
   \put(121,5){\line(0,1){87}}
     \put(33,109){\line(1,0){95}}
      \put(95,-2){$z_{*_0}$}
        \put(128,-5){$z_{h}$}
\put(170,-3) {\includegraphics[width=4.3cm]{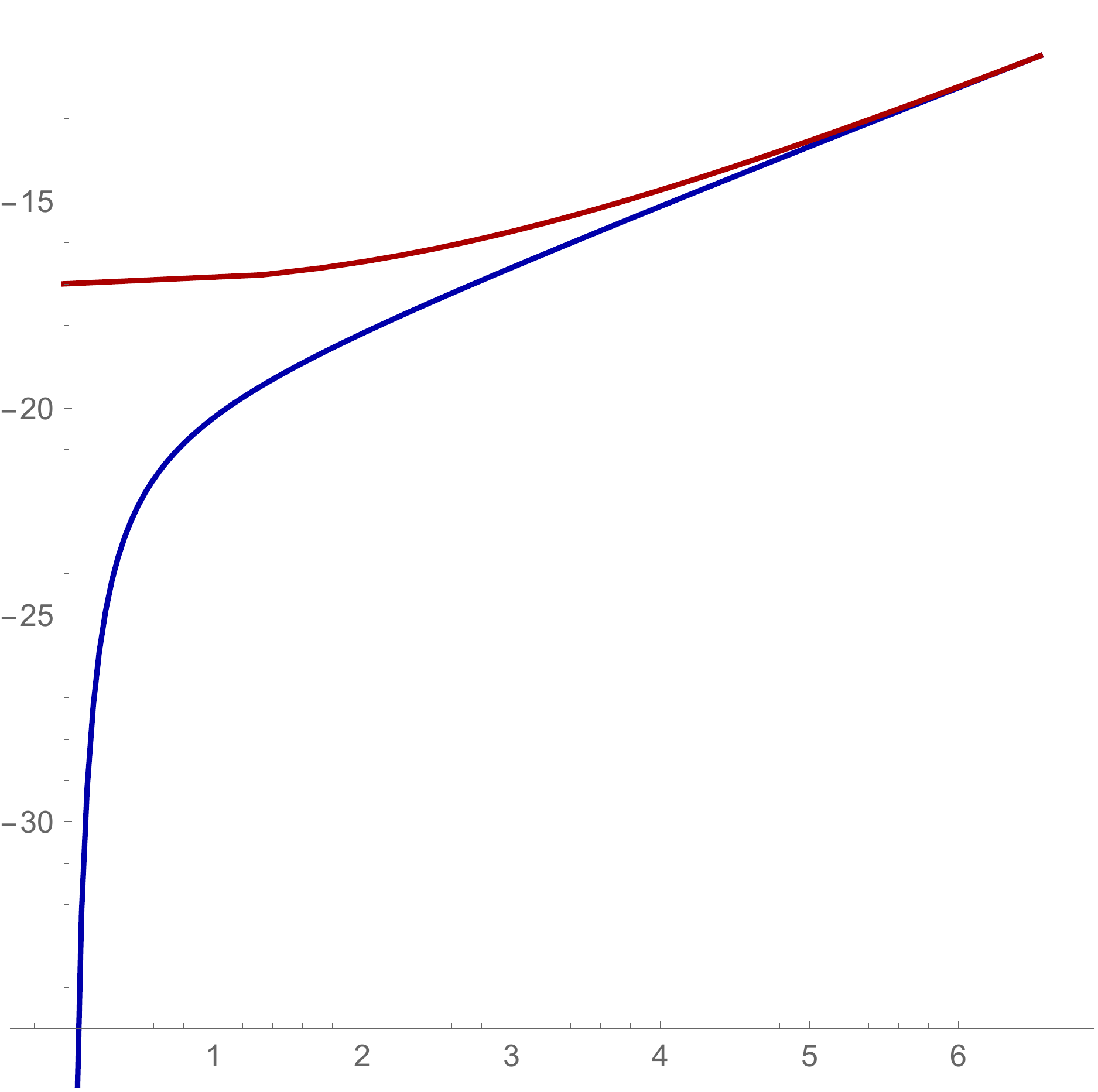}}
 \put(170,135){$E_{x} $}
   \put(280,5){\line(0,1){112}}
    \put(177,110){\line(1,0){117}}
   \put(185,115){$E_{x_0} $}
     \put(276,-7){$\ell_0$}
       \put(290,-2){$\ell$}
\put(0,-20){a)}
  \put(150,-20){b)}
   \put(290,-20){c)}
 \end{picture}\\
 $\,$
 \caption{
 Above the critical point (deconfinment): a) $L=L(z_*)$, b) $E_x=E_x(z_*)$ and c)$E_x=E_x(\ell)$. 
 Here we put $\alpha^{'}=1/\pi$}
\label{fig:L-zs-above}
\end{figure}

\item There are two extremal points in the interval $0<z<z_{h_{i_0}}$, $z_{min}=z_0$ and $z_{max}=z_0'$
and the potential is a decreasing function only on the intervals
\be\label{reg-conf}
0<z<z_{min} \,\,\,\,{\mbox {and }}\,\,\,\,z_{max}<z<z_{h_{i_0}},\ee
so we can guarantee that
$
 V(z)/V(z_*)<1$ only  in the region \eqref{reg-conf}. 
 
This case corresponds to  the confinement phase and the dependence of $L$ on $z_*$ is presented in Fig.\ref{fig:L-zs-below}.a). Suppose that $V(z)$ has a local minimum at $z=z_0$. In this case the potential $V_x(z)$ is a decreasing function from 
$z=0$ up to $z=z_0$. The first branch $L=L(z)$ is shown by the blue line in the plot in Fig.\ref{fig:L-zs-below}.a). and $L(z)\to \infty $ when $z\to z_0-0$. For $z_0<z<z_0'$ $L=L(z)$ gets  an imaginary part and at $z=z'_0$ starts the second branch. 
For the second branch $E_2<E(\ell)<E^\prime_2$.

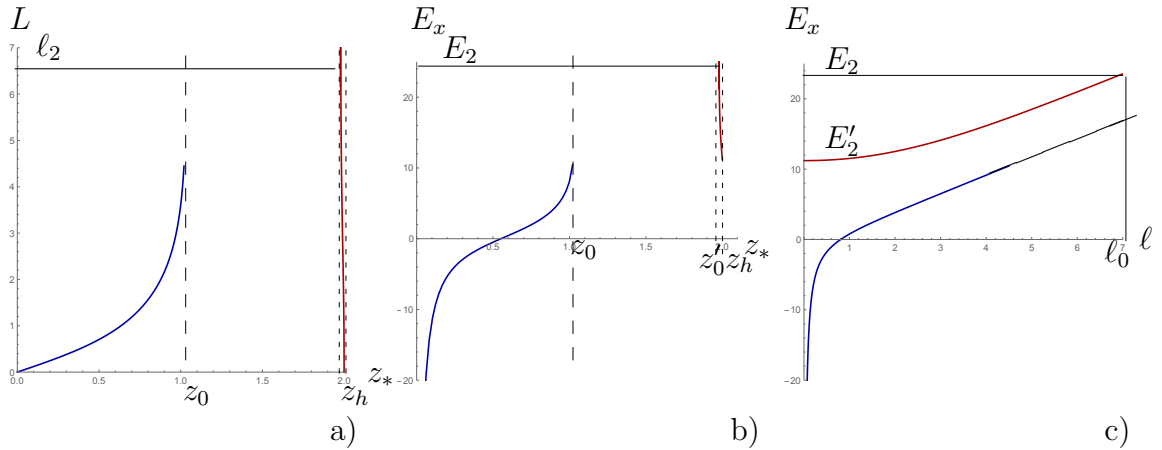
\begin{figure}[h!]
\centering\begin{picture}(185,150)
\put(-120,0){\includegraphics[width=4.5cm]{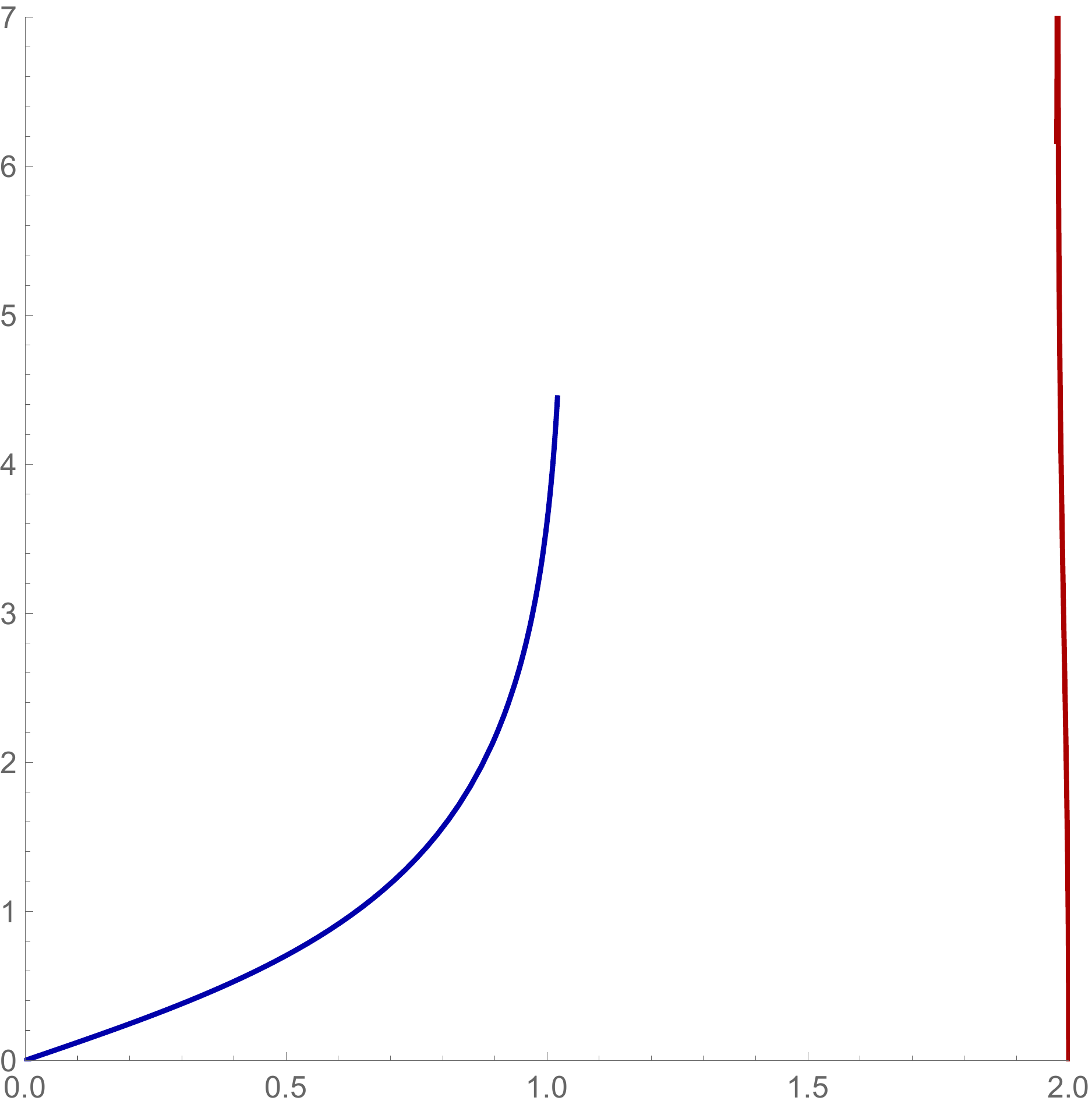}}
 \put(-120,135){$L$}
 \put(-118,120){\line(1,0){120}}
 \linethickness{0.1mm}
 \multiput(-54,10)(0,10){12}{\line(0,1){5}}
 \multiput(3.5,5)(0,5){25}{\line(0,1){2}}
 \multiput(6,5)(0,5){25}{\line(0,1){2}}
 \put(10,2){ $z_*$}
 \put(-55,-5){$z_{0}$}
  \put(4,-5){$z_{h}$}
     \put(-110,125){$\ell _{2}$}
\put(25,0){\includegraphics[width=4.5cm]{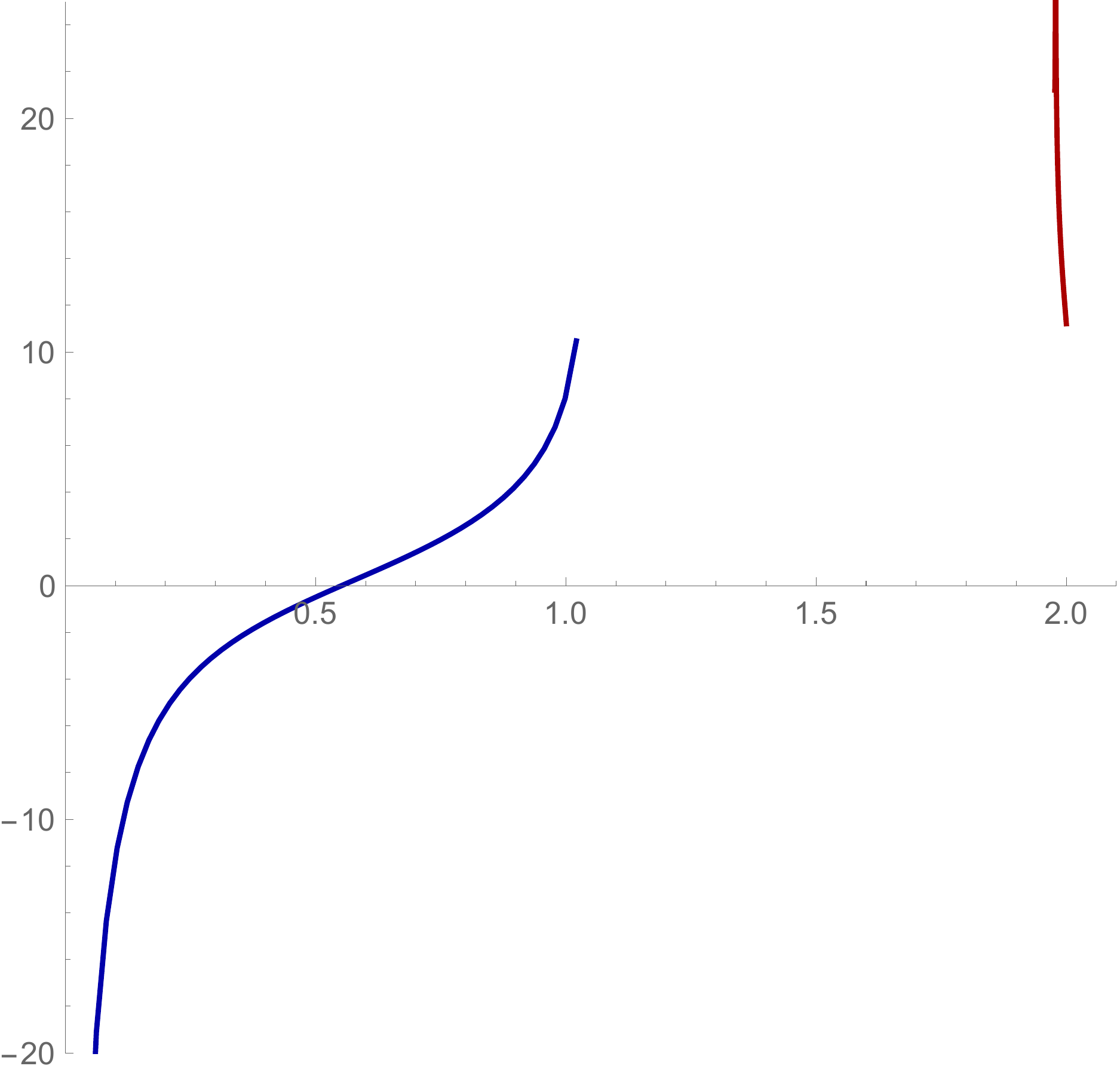}}
 \put(30,135){$E_{x} $}
  \put(42,124){$E_{2} $}   
    \linethickness{0.1mm}
 \multiput(91,10)(0,10){12}{\line(0,1){5}}
 \multiput(144.5,50)(0,5){15}{\line(0,1){2}}
 \multiput(147,50)(0,5){15}{\line(0,1){2}}
   \put(33,121){\line(1,0){113}}
      \put(90,50){$z_{0}$}
       \put(138,45){$z'_{0}$}
        \put(148,45){$z_{h}$}
         \put(155,50){$z_*$}
\put(170,0) {\includegraphics[width=4.5cm]{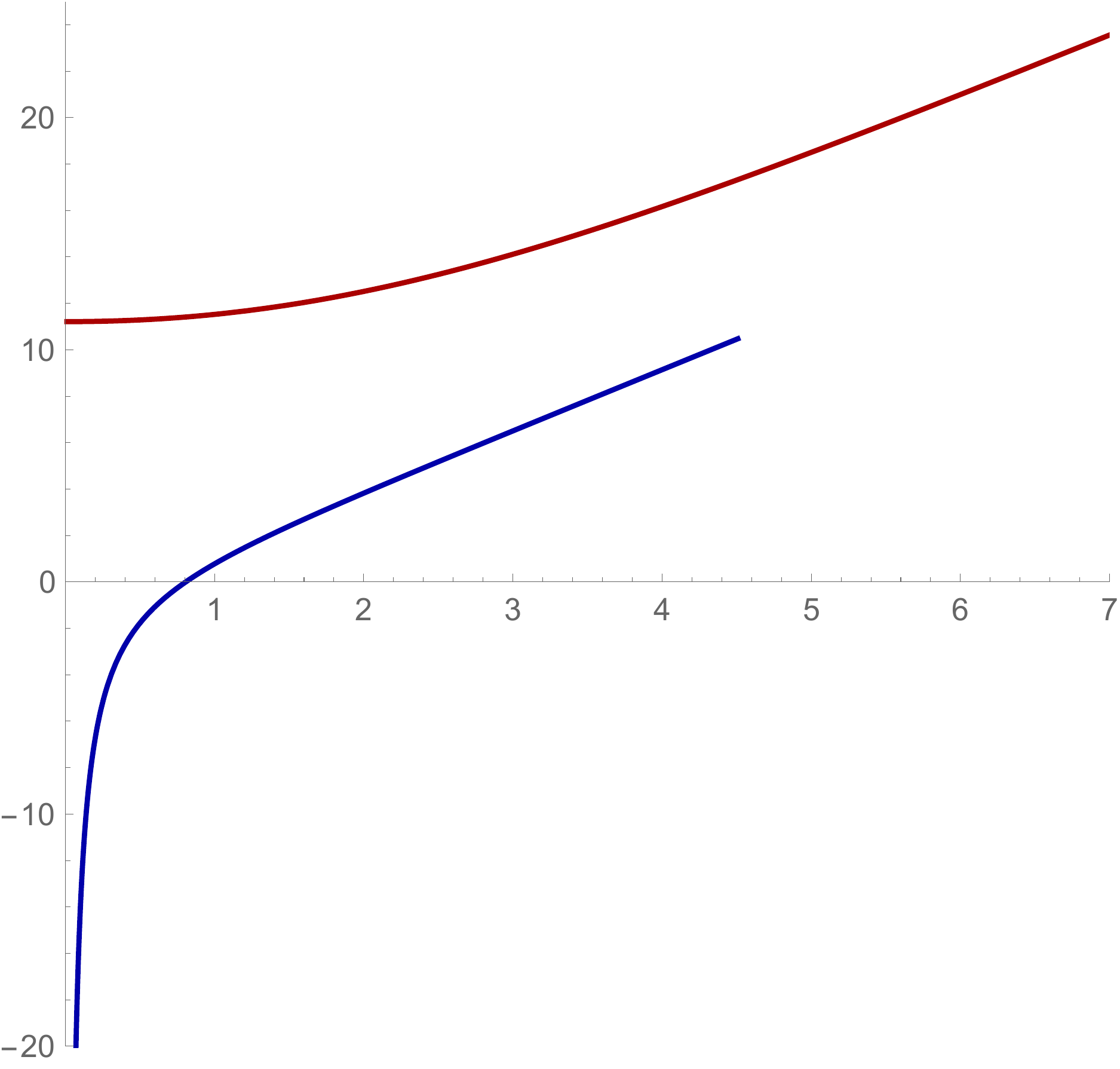}}
 \put(170,135){$E_{x} $}
   \put(298,55){\line(0,1){62}}
    \put(177,117.5){\line(1,0){120}}
      \put(247,80.5){\line(5,2){55}}%
   \put(185,120){$E_{2} $}
      \put(185,90){$E'_{2} $}
     \put(290,47){$\ell_0$}
       \put(303,52){$\ell$}
\put(0,-20){a)}
  \put(150,-20){b)}
   \put(290,-20){c)}
 \end{picture}\\
 $\,$
 \caption{
  Below the critical point (confinment): a) $L=L(z_*)$, b) $E_x=E_x(z_*)$ (note that in this case we make a different subtraction) and c) $E_x=E_x(\ell)$ 
 }
\label{fig:L-zs-below}
\end{figure}
\end{itemize}

\begin{figure}[h!]
\centering
\includegraphics[scale=0.15]{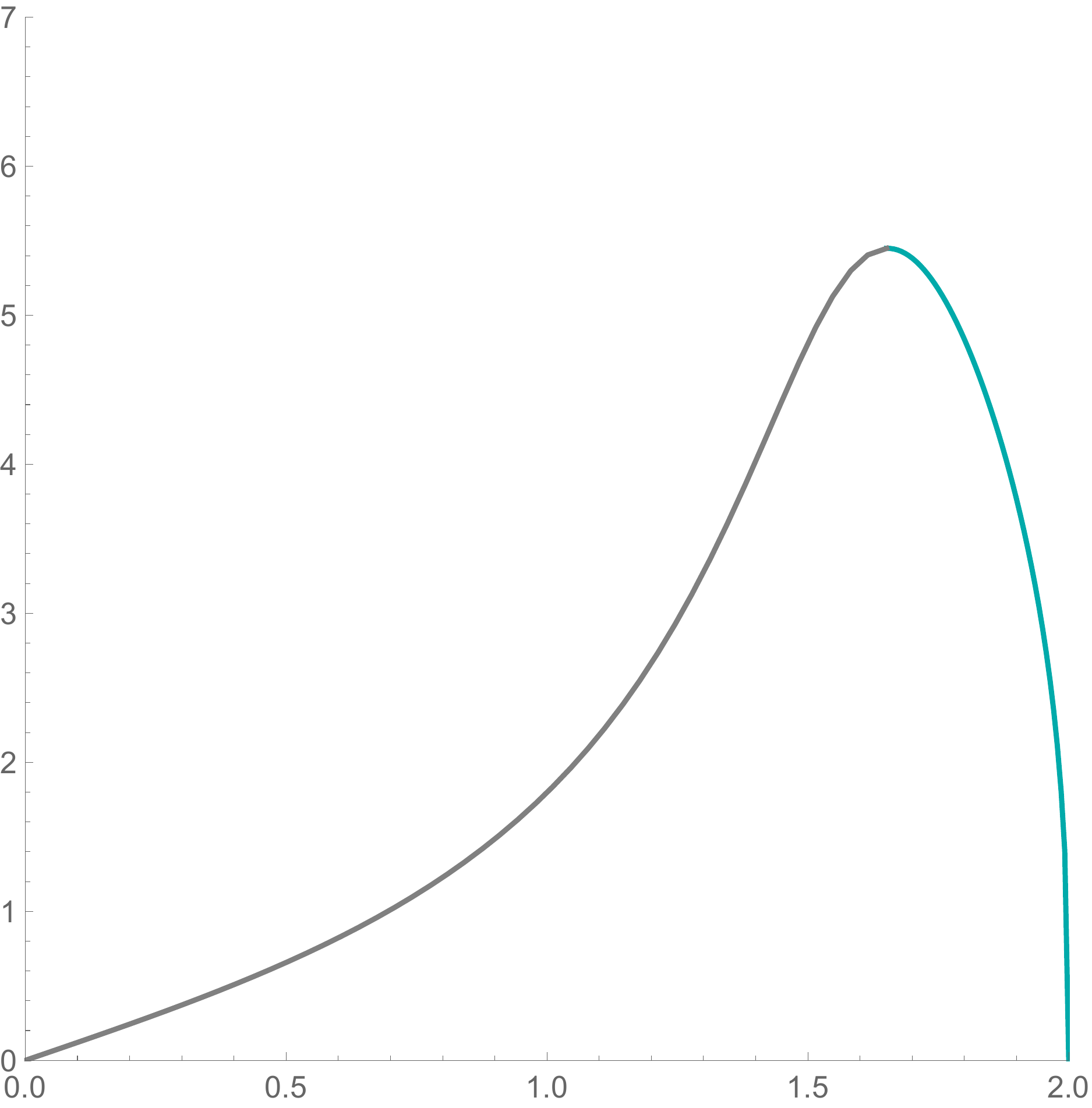}$\,\,\,a\,\,$$\,\,\,\,\,$
\includegraphics[scale=0.15]{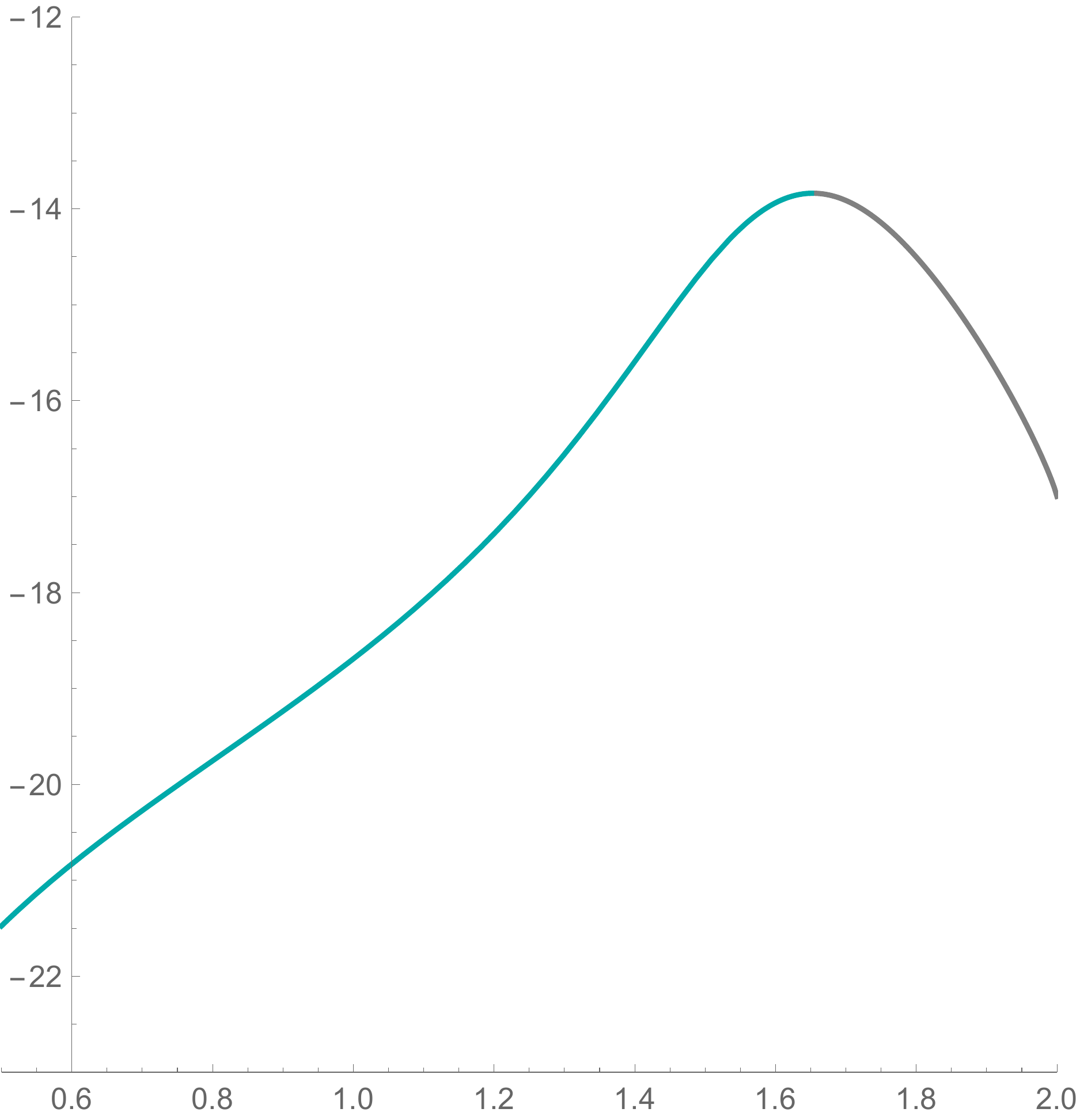}$\,\,\,\,b\,$$\,\,\,\,\,$
\includegraphics[scale=0.15]{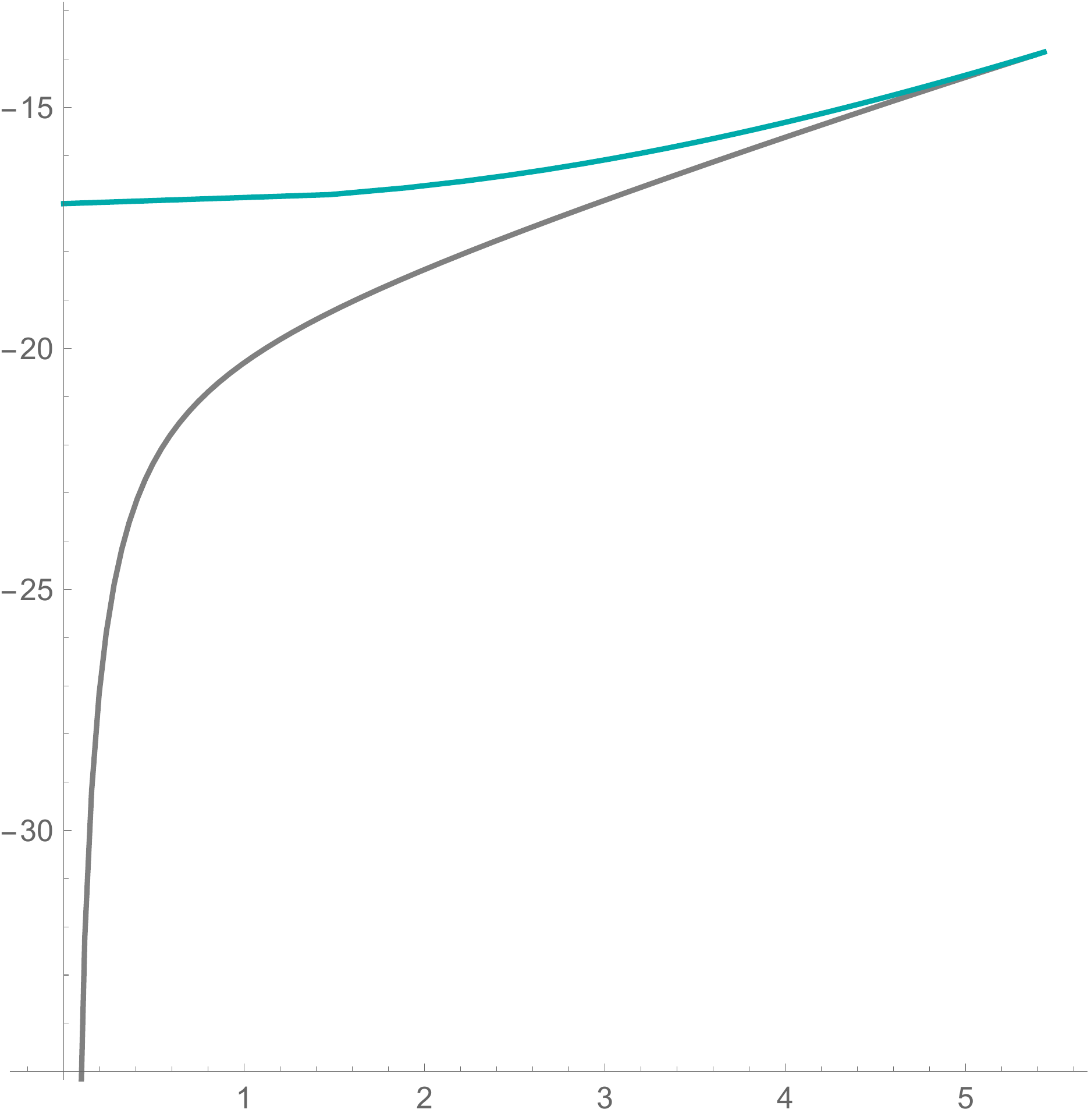}$\,\,\,\,c$\\$\,$\\
\includegraphics[scale=0.15]{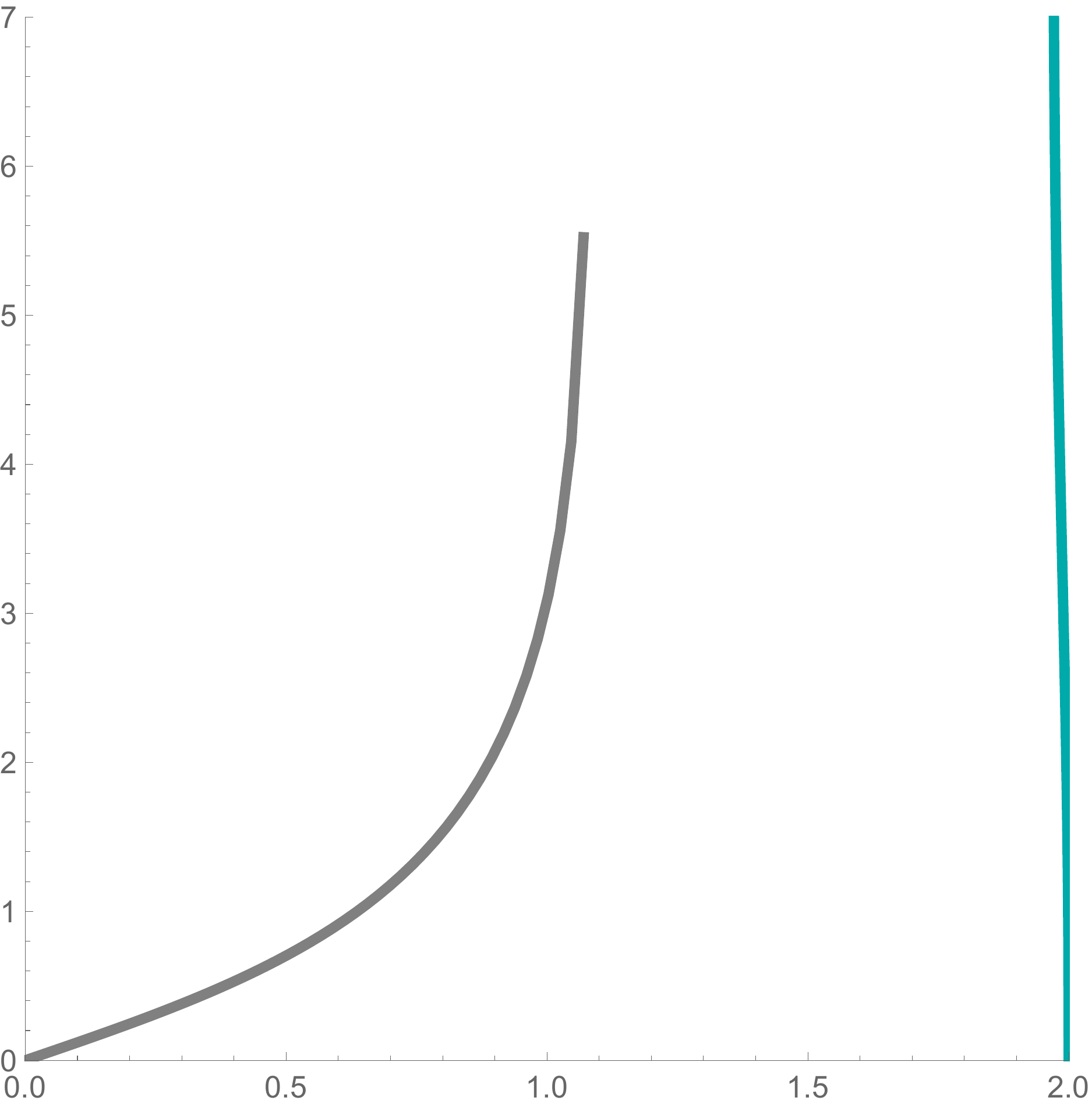}$\,\,\,d\,\,$$\,\,\,\,\,$
\includegraphics[scale=0.15]{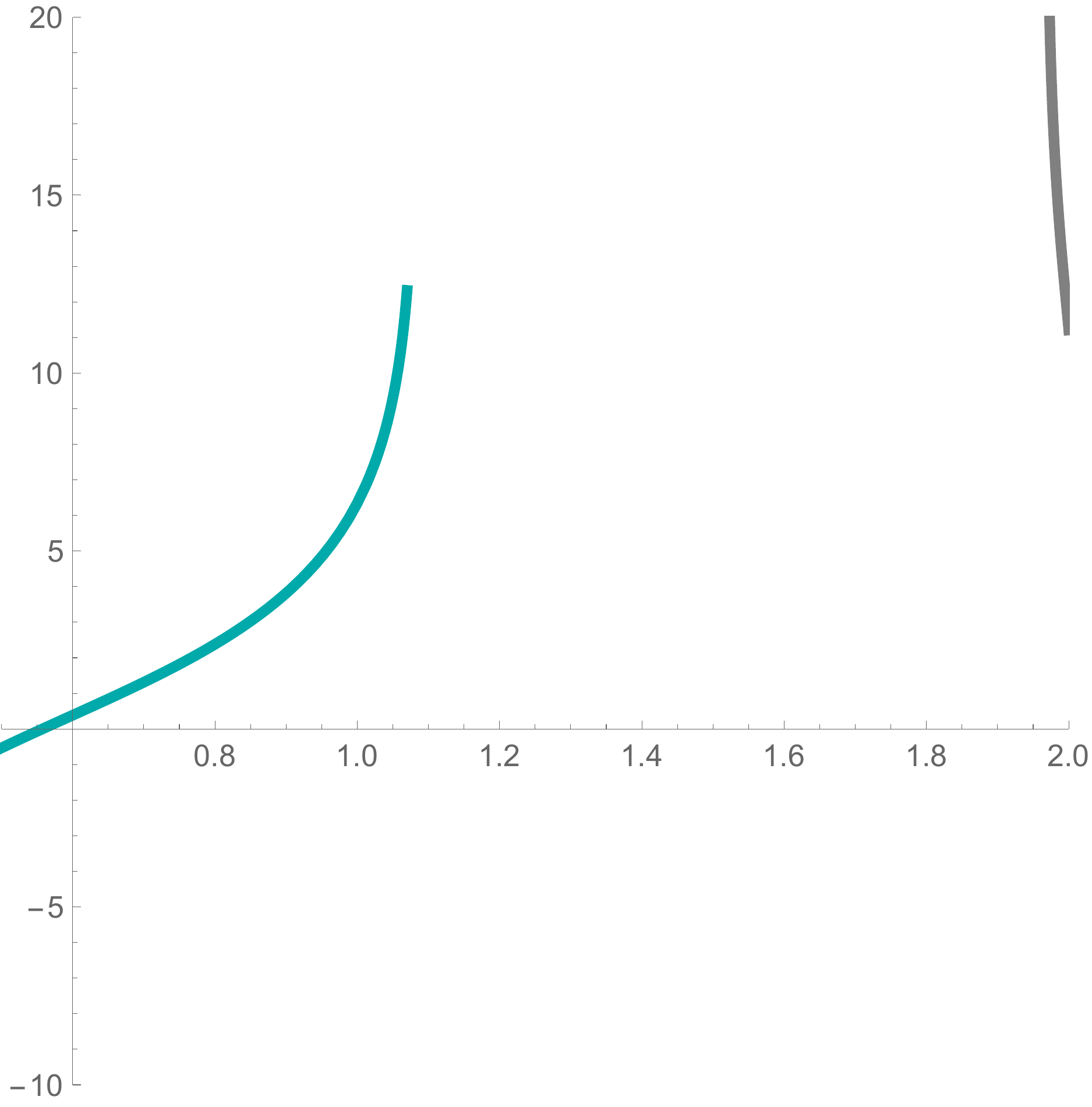}$\,\,\,\,e\,$$\,\,\,\,\,$
\includegraphics[scale=0.15]{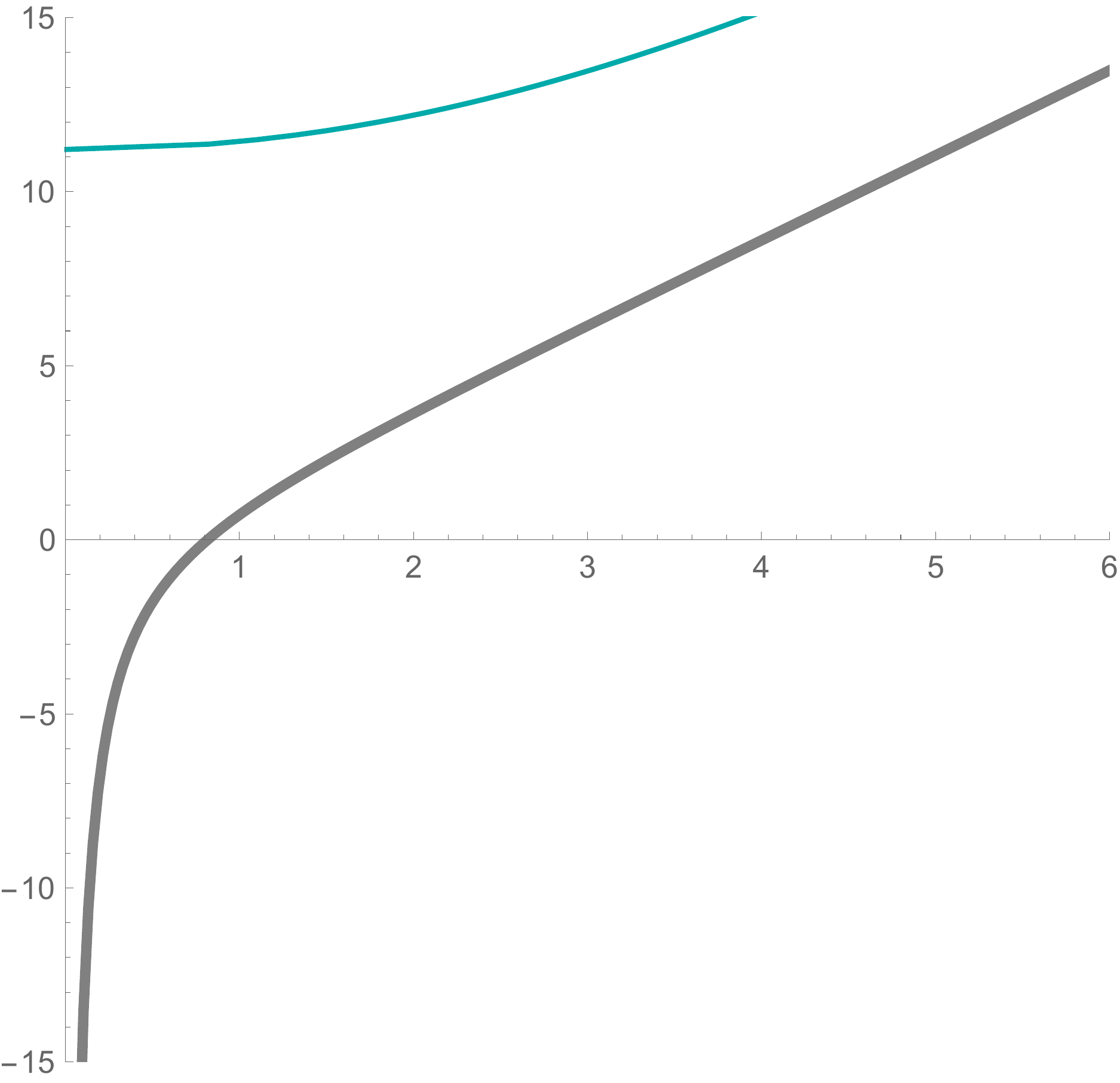}$\,\,\,\,f$\\
 \caption{
 The same quantities as in Fig.\ref{fig:L-zs-above} and \ref{fig:L-zs-below} at the same values of parameters, except that now we deal with the anisotropic case, $\nu=4$. a),b) and c) above the critical point (deconfinment): a) $L=L(z_*)$, b) $W=W(z_*)$ and c) $W=W(L)$. 
 d), e) and f) below the critical point (confinment): d) $L=L(z_*)$, e) $W=W(z_*)$ (note that in this case we make a different subtraction)   and f) $W=W(L)$.
 }
\label{fig:L1-zs4}
\end{figure}

In Fig.\ref{fig:L1-zs4} we present the same quantities as in Fig.\ref{fig:L-zs-above}  and 
Fig.\ref{fig:L-zs-below} at the same parameters except that 
in Fig.\ref{fig:L-zs-above}  and  Fig.\ref{fig:L-zs-below} $\nu=1$ (isotropic case) and in 
Fig.\ref{fig:L1-zs4} $\nu=4$ (anisotropic case, longitudinal orientation of the Wilson loop). 
The confinement solution is realized on the first branch discussed above.
Let us estimate the string tension for this configuration. Suppose that 
\be
V'(z)|_{z=z_0}=0, \,\,\,\,\,\,V^{\prime \prime}(z_0)>0.\ee
Near $z=z_0$  we have $V^{\prime \prime}(z_0)>0$ and
\bea
\sqrt{\frac{V(z)}{V(z_0)}^2-1}&=&\sqrt{\frac{V^{\prime \prime}(z_0)}{V(z_0)}}\,(z-z_0)+{\cal O}(z-z_0)^2\\
\label{L1bmmmm}
L_{1} &\underset{z_*\sim z_0}\sim &
-2\sqrt{\frac{V(z_0)}{f(z_0)V^{\prime \prime}(z_0)}}\ln (z_0-z_*)\eea

We can also calculate the energy near $z_*=z_0$ using \eqref{WLxb-mm}.  We have

\begin{eqnarray}\label{NG1Em}
S_{xt}  &\underset{z_*\sim z_0}\sim&
-\frac{T}{2\pi \alpha^{'}} \frac{ b(z_0)} {z_0^2}
\frac{1}{\sqrt{\frac{V^{\prime \prime}(z_0)}{V^(z_0)}}}\ln(z_0-z_*)
 \end{eqnarray}
Therefore 
we get
\bea\nn
S_{xt}&\underset{L\to \infty}{\sim}&
 \frac{T}{\pi \alpha^{'}}
 \frac{L_{1}}{2} \frac{ b(z_0)\sqrt{f(z_0)}} {z_0^2}\eea
 and
 \bea\label{sigma-x} \sigma_{x}
 &=& \frac{V_x(z_0)}{2\pi \alpha^{'}} 
\eea

\subsubsection{Energy between quarks located along y-direction in  the  quark confinement background \eqref{ds-aniz-con-chem} }
\begin{figure}[h!]
\centering
\begin{picture}(185,125)
\put(0,0){\includegraphics[scale=0.35]{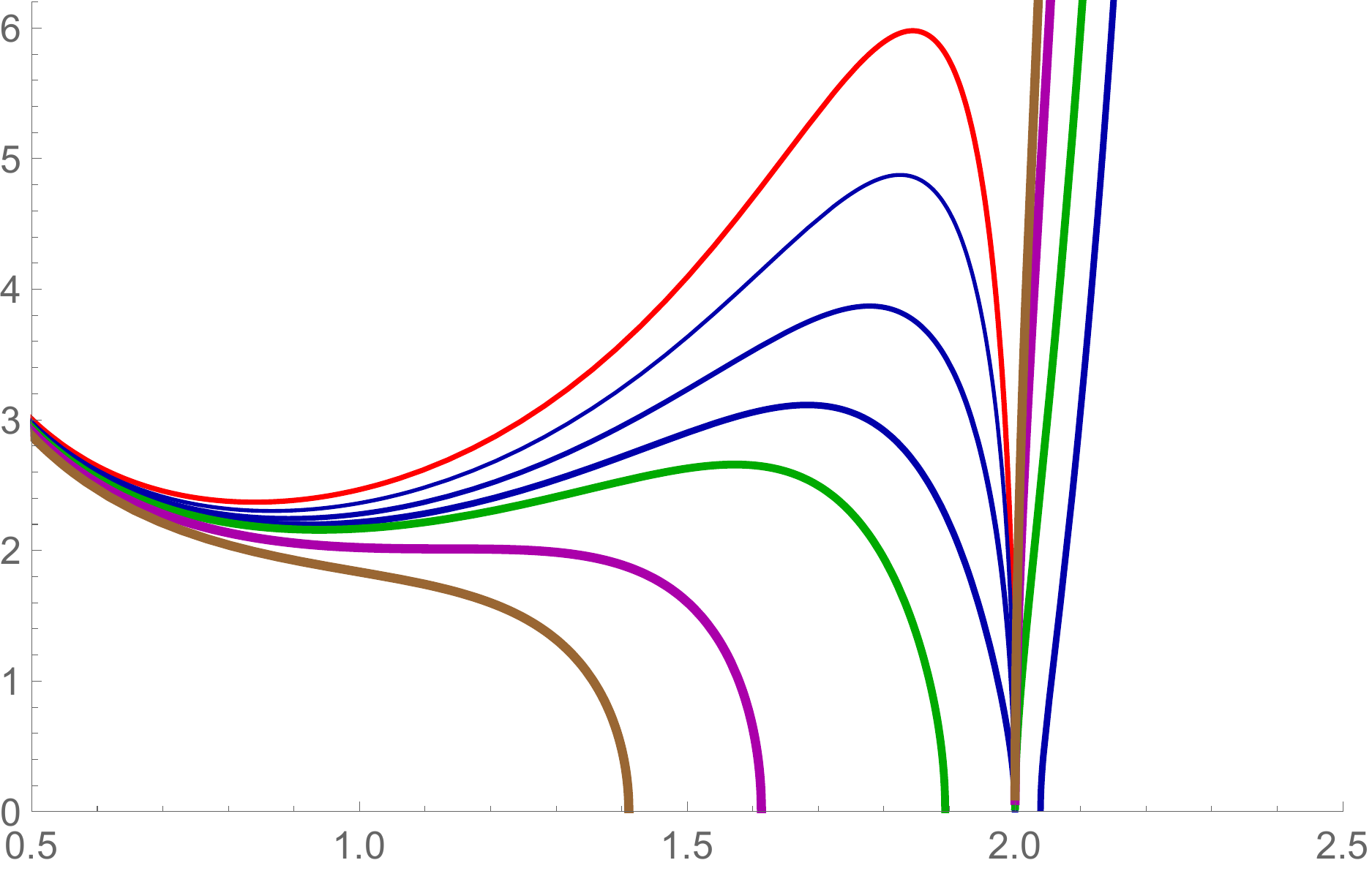}}
 \put(0,130){$V_y$}
  \put(195,8){$z$}
\end{picture}
 \caption{
 The potential $V_y(z)=V(z, c, q,\nu,z_h)$ given by  \eqref{Vyz} with $b(z)$
 given by \eqref{b-AZ}  with $c=2$ and $f(z)=f(z,q,\nu,z_h)$ given by \eqref{f-q-zh-nu} with $\nu=4$  and $z_h=2$. The red line 
 shows $V_y(z)$ for  $q=0$. We see that the corresponding potential  has one minimum and one maximum.
 The blue lines 
 show $V_y(z)$ for    $q=0.15,0.2$ and  $q=0.23$. The corresponding potentials have 
 minimum and maximum.  The green line shows $V_y(z)$ for    $q=0.25$ and second zeros of the corresponding function $f(z)$
is at $z_{h_{2}}=1.89<z_h$. The magenta line 
 shows $V(z)$ for    $q=0.3$. The corresponding potential   has no
 minimum and maximum and  the second zero of the corresponding function $f(z)=f(z,q,1,z_h)$
 is at $z_{h_2}=1.61$. The brown line
 shows $V_y(z)$ for  $q=0.35$  and the corresponding potential has no
 minimum and maximum  on $0<z<z_{h_2}$, where $z_{h_2}=1.40$ is the second zero of the corresponding functions $f(z)$.  The critical charge $q_{cr}|_{c=2,z_h=2}=0.3$
  }\label{fig:pot-VV-z-q-c2-nu4}
\end{figure}

  The Nambu-Goto action $S_{yt}$ of the string  hanging from 
 the contour $C$, being the rectangular with  sites along the time direction and one of transversal directions in the bulk,
 is
\bea\label{S-ty}
S_{y t}&=&\frac{Ti}{2\pi\alpha^{\prime}}\int \frac{b(z)} {z^2} \sqrt{z^{2-2/\nu}f(z)+z^{\prime 2}}\,dx,
\eea
this formula has an explicit dependence on the anisotropy and is in agreement with the general formula obtained in \cite{DiGi}.  The corresponding "potential" is, see Fig.\ref{fig:pot-VV-z-q-c2-nu4},
\be\label{Vyz}
V_y(z)= \frac{b(z)\sqrt{f(z)}}{z^{1/\nu+1}}
\ee
and  for $L_y$  we have
 
  \be
  \label{Lybm}
 L_{y}   = 2\int^{z_*}_{0} \frac { z^{1/\nu-1}\,dz}{\sqrt{f(z) }
 \sqrt{\frac{V^2(z)}{V^2(z_*)}-1}}.
  \ee

To get for the energy of the string
 between two quarks  stretched in  the $y$-direction we  perform the subtraction of the two quark mass that  corresponds the subtraction of the action on the sheet starting at the boundary $z=0$
and ending at the horizon 

     \bea
     \label{Ey}
\pi \alpha^{'}E_{y} &= &\int^{z_{*}}_{0} \frac{dz }{z^2}\left[\frac{b(z)V_y(z)}{\sqrt{V_y^2(z) -V_y^2(z_*)}}
-1
\right]
-\frac{1}{z_*}+
m^{\frac{\nu }{2 \nu +2}} e^{c\, m^{-\frac{2 \nu }{2 \nu
   +2}}}-\sqrt{\pi c} \,\text{erfi}\left(\sqrt{c}\,
   m^{-\frac{\nu }{2 \nu +2}}\right).
\label{WLS2}
\eea
   
   Note that the expression for $E_y$, \eqref{Ey},  in term of the potential $V$ is the same as $E_x$, \eqref{Ey}, 
   the only difference is in the form of the potentials. Supposing that $V_y(z)$ has a minimum at $z=z_0$, we can estimate the contributions near this minimum, we 
   have
\be\label{WLybmin}
\pi \alpha^{'}E_{y}\underset{z_*\sim z_0}\sim    \frac{\ln(z_0-z_*)} {z_0^{2}} \frac{ b(z_0)}{\sqrt{\frac{V"(z_0)}{V(z_0)}}},
\,\,\,\,\,\,
 L_{y}  \underset{z_*\sim z_0}\sim  \frac { z_0^{1/\nu-1}\ln (z_0-z_*)}{\sqrt{f(z_0) }}
\frac{ 1}{\sqrt{\frac{V^{\prime\prime}(z_0)}{V(z_0)}}}
  \ee 
and we get the  answer similar to  \eqref{sigma-x},
 \be \label{sigma-y}\sigma_{y}
 =\frac{V_y(z_0)}{2\pi \alpha^{'}} .
\ee
\begin{figure}[h!]
\centering
\includegraphics[scale=0.15]{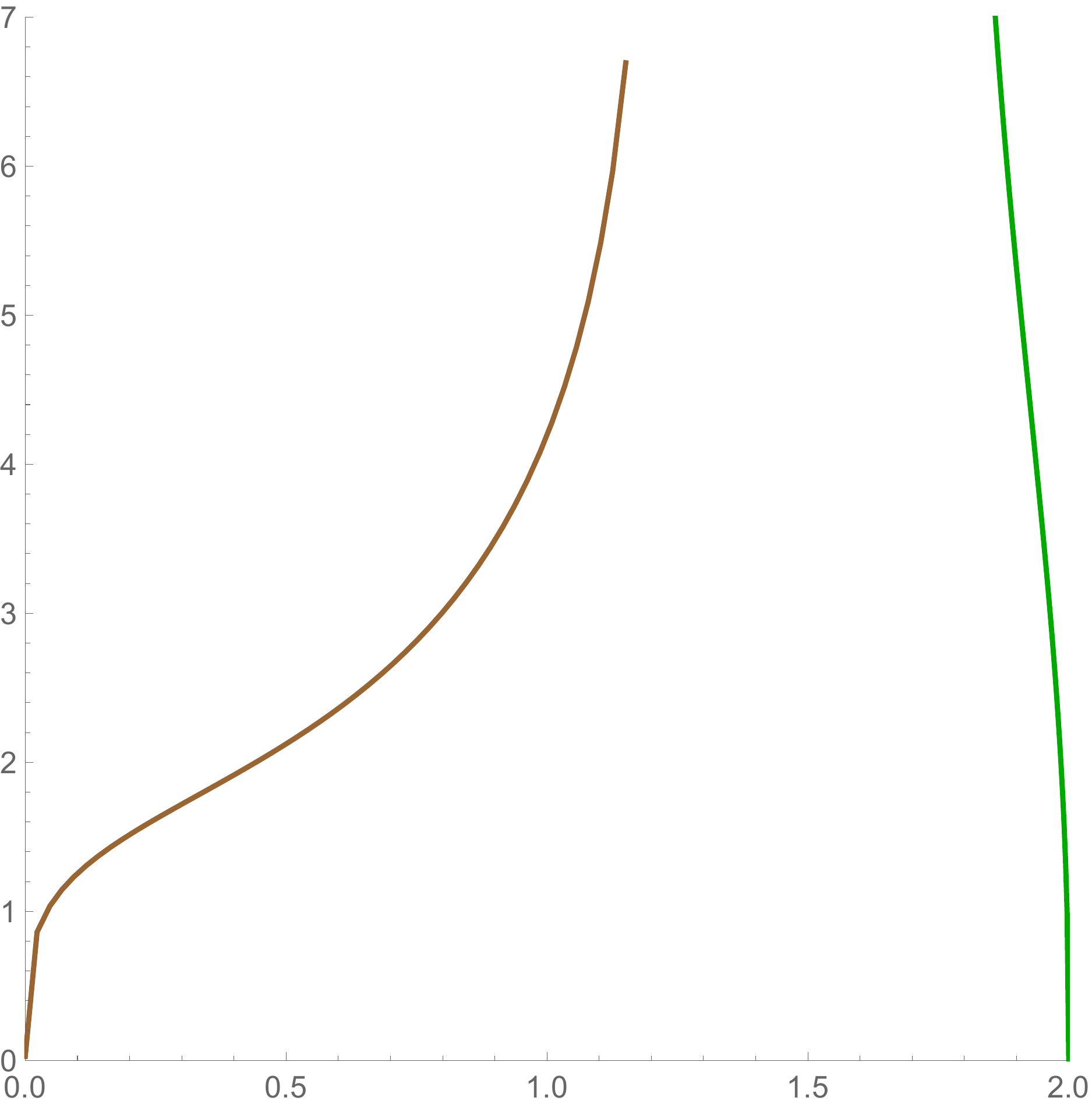}$\,\,\,a\,\,$$\,\,\,\,\,$
\includegraphics[scale=0.15]{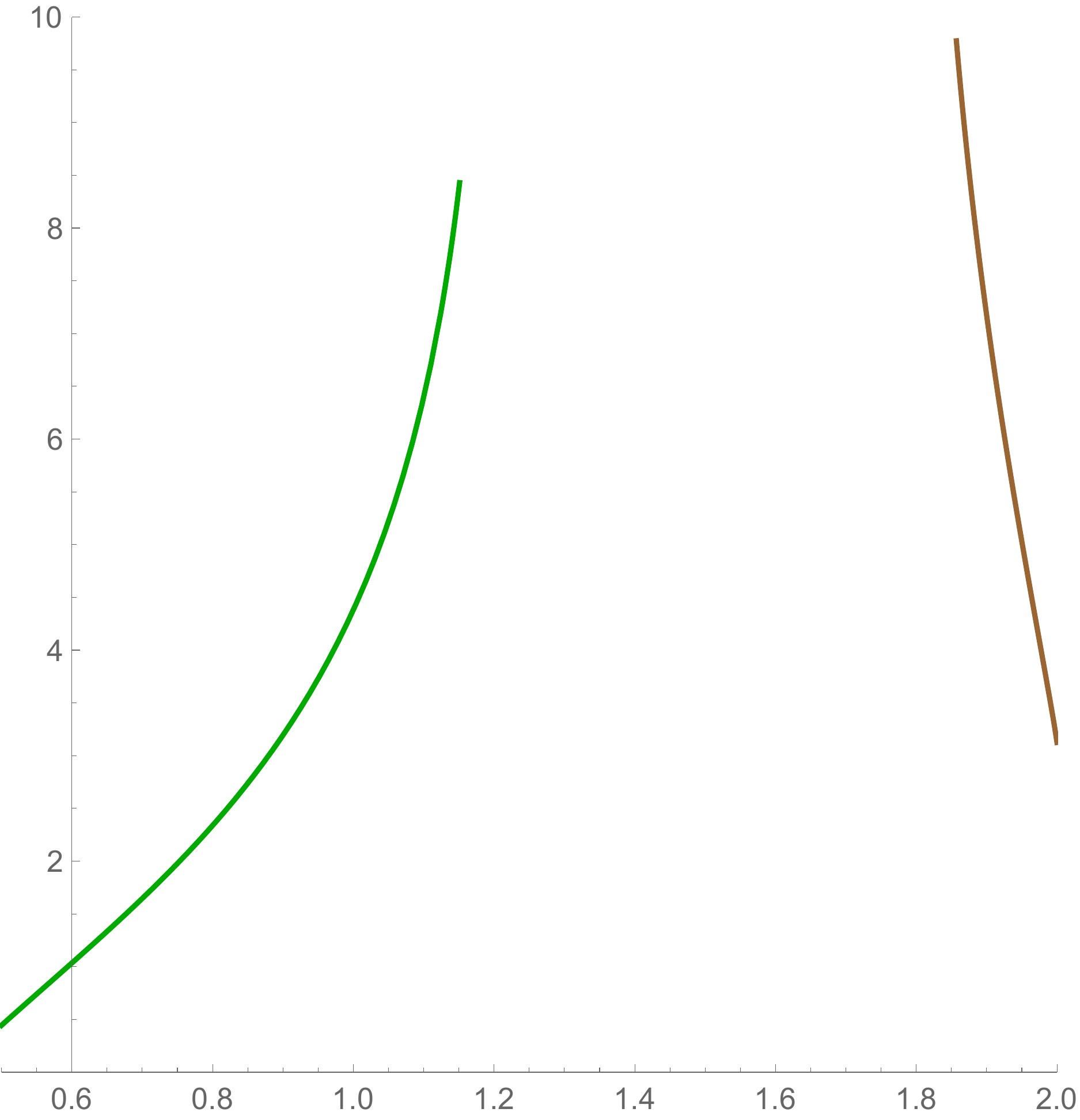}$\,\,\,\,b\,$$\,\,\,\,\,$
\includegraphics[scale=0.15]{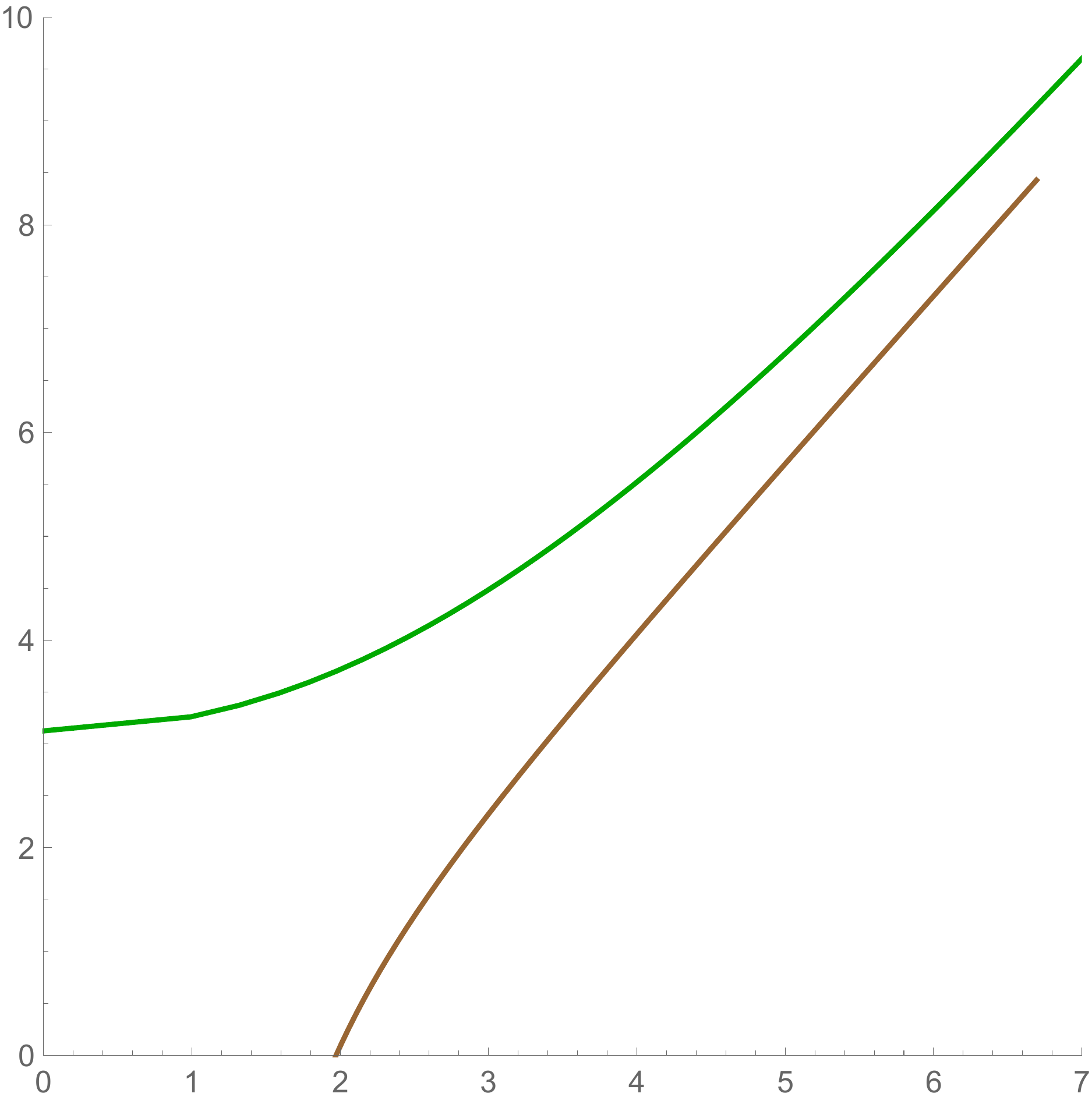}$\,\,\,\,c$\\$\,$\\
\includegraphics[scale=0.15]{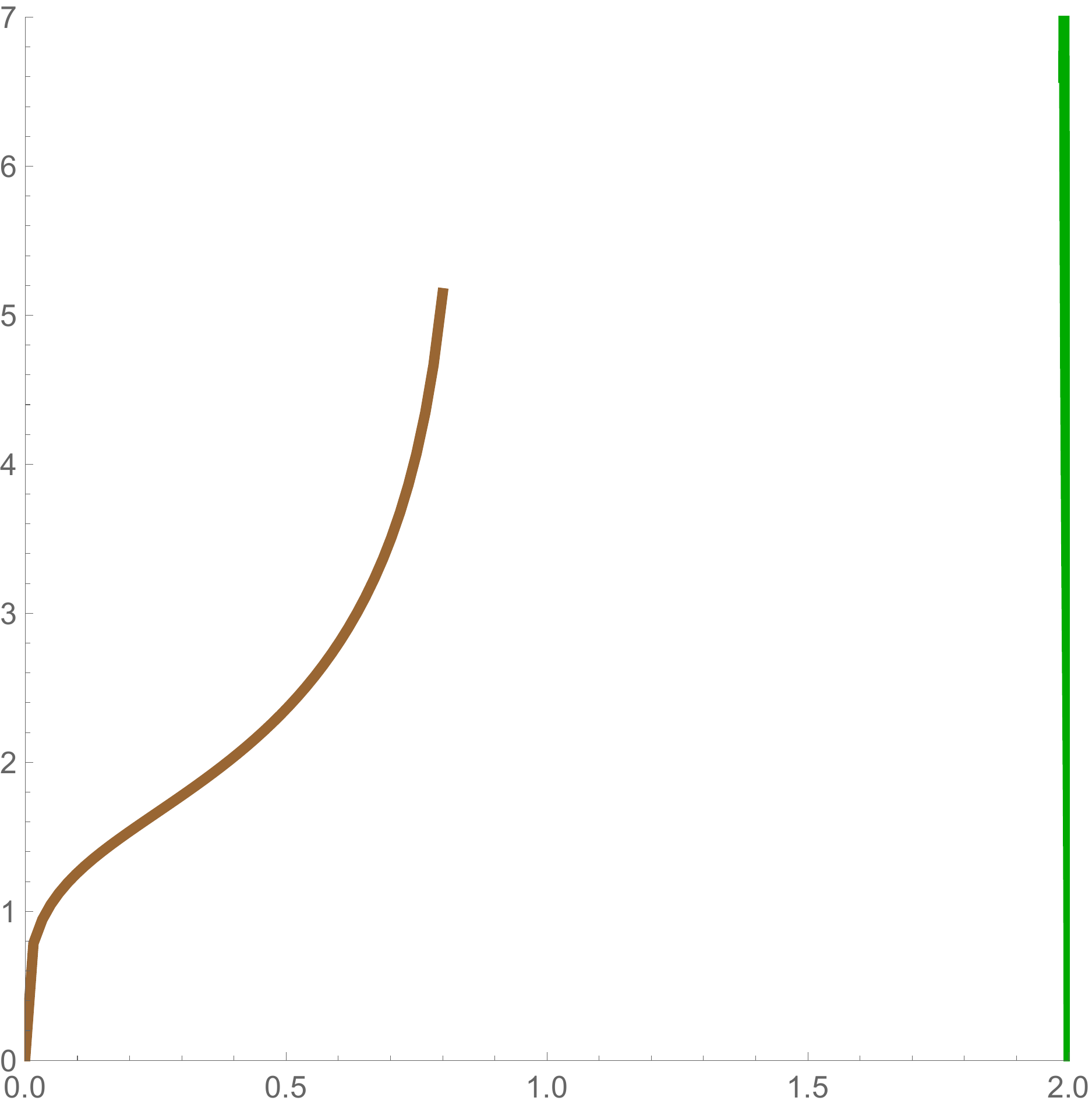}$\,\,\,d\,\,$$\,\,\,\,\,$
\includegraphics[scale=0.15]{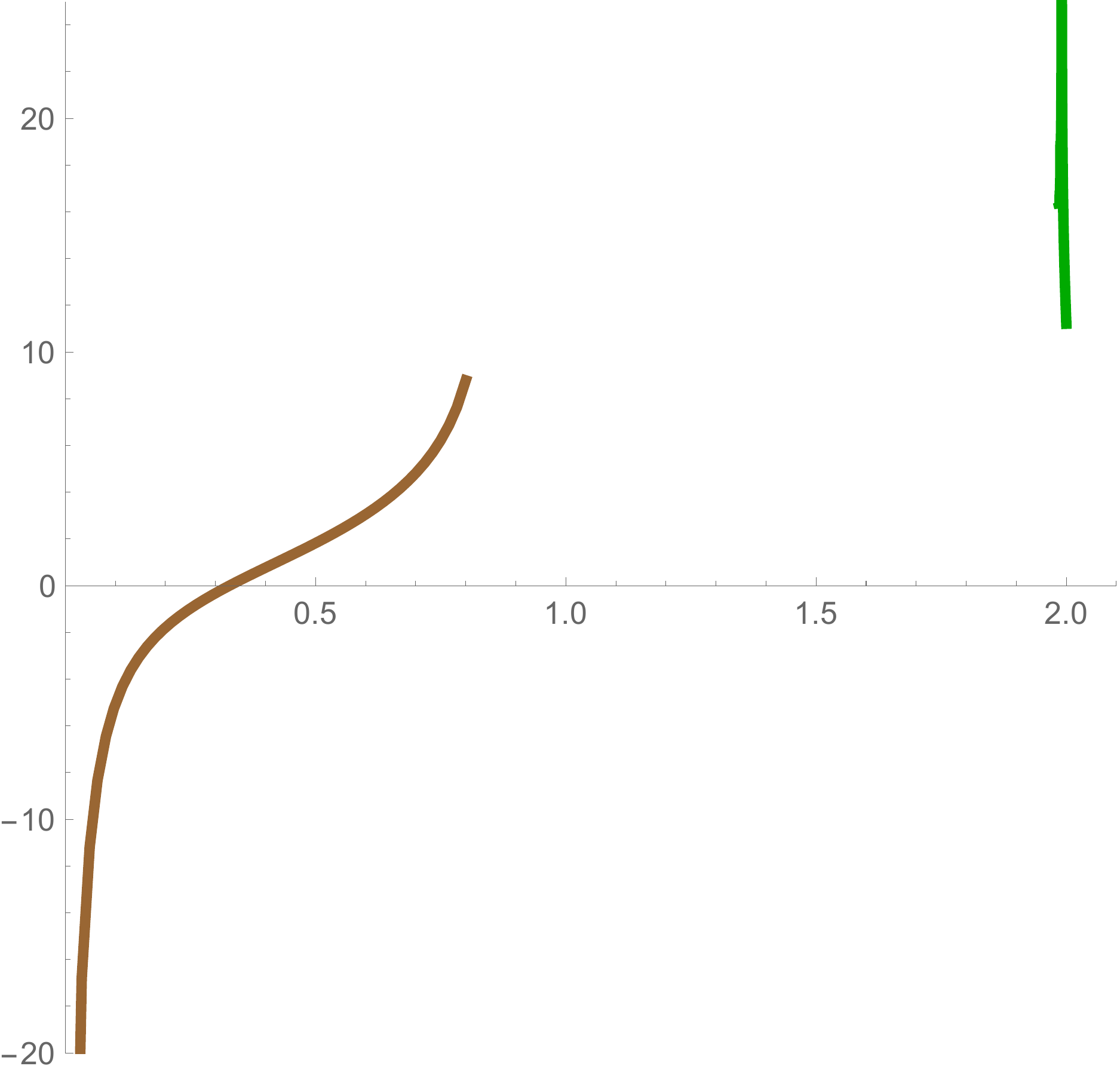}$\,\,\,\,e\,$$\,\,\,\,\,$
\includegraphics[scale=0.15]{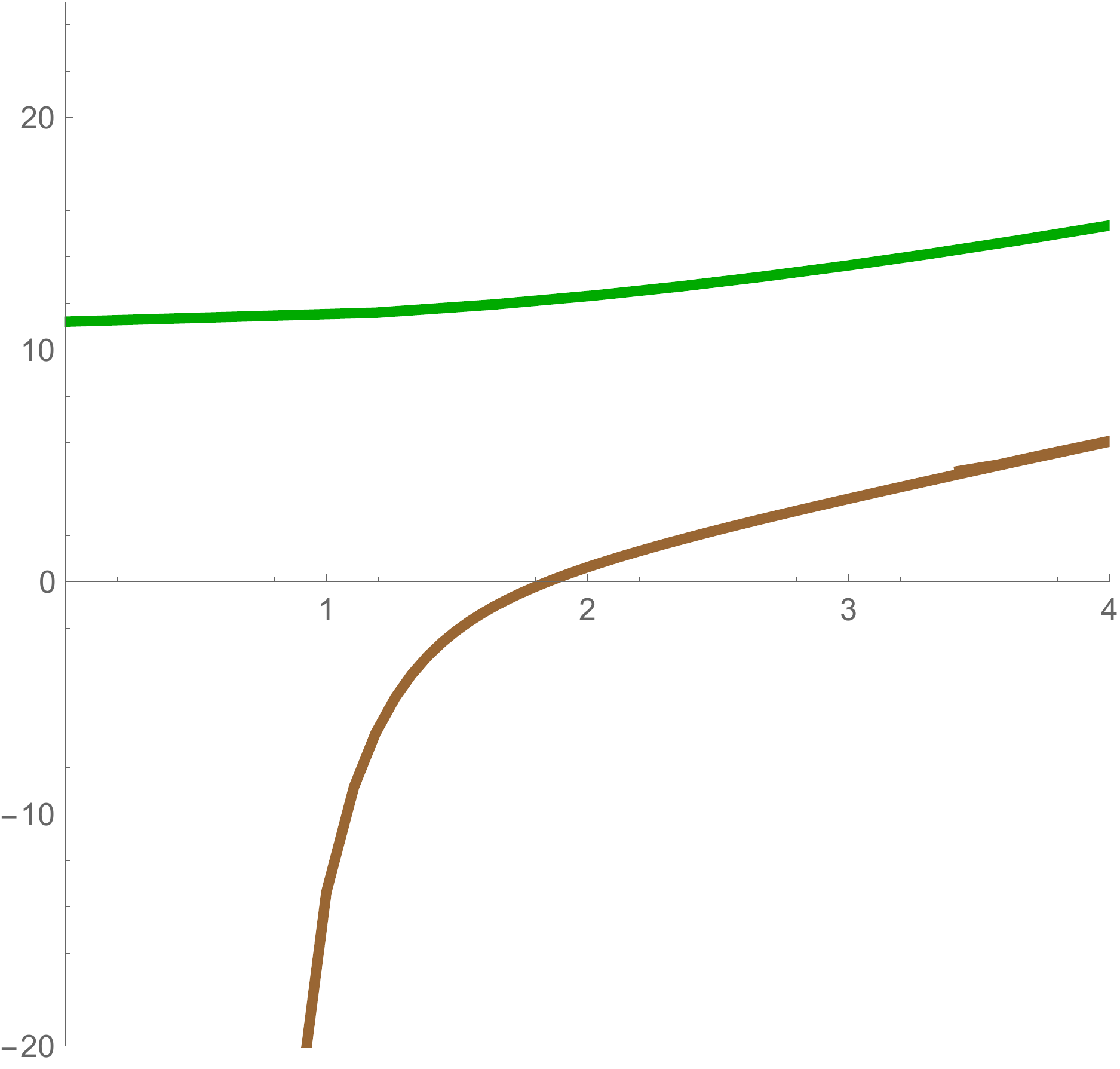}$\,\,\,\,f$
 \caption{
 The same quantities at the same parameters as in Fig.\ref{fig:L1-zs4} except that now we deal with the transversal orientation of the Wilson loop, $\nu=4$. We see that both $c=1.2$ and $c=2$ correspond to the confinement phase:  a) and d) $L=L(z_s)$, b) and e) $W=W(z_s)$ (note that in this case we make a different subtraction), c) and 
 f)  $W=W(L)$ .
 }
\label{fig:L2-zs}
\end{figure}

In Fig.\ref{fig:L2-zs} we present the same quantities as in Fig.\ref{fig:L-zs-above}  and 
Fig.\ref{fig:L-zs-below} at the same parameters except that 
in Fig.\ref{fig:L-zs-above}  and  Fig.\ref{fig:L-zs-below} $\nu=1$ (isotropic case) and in 
Fig.\ref{fig:L2-zs} $\nu=4$ (anisotropic case, transversal orientation of the Wilson loop). We see that for the transversal orientation at the same parameters as for the isotropic case we can get different solutions: for the isotropic case, as well as for the anisotropic longitudinal case, $c=1.2$ corresponds to 
the deconfined phase, meanwhile this value of $c$ in the anisotropic transversal case  corresponds to
the confined phase. This observation indicates that the phase diagram should be essential depends on the orientation in the case of the anisotropic background.

 \subsection{Holographic anisotropic QCD phase diagrams}
\begin{figure}\centering\begin{picture}(185,130)
\put(0,0){\includegraphics[width=7cm]{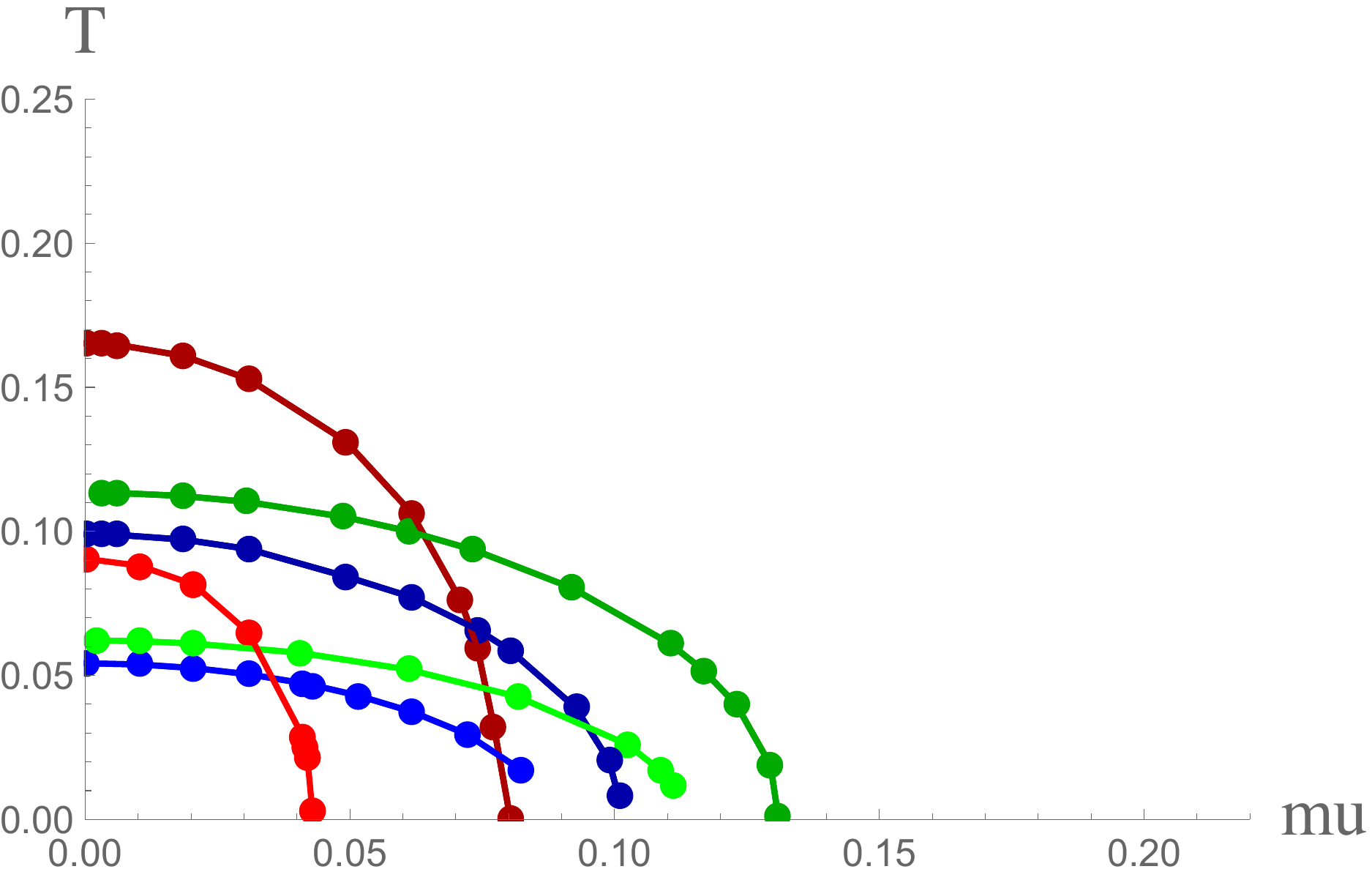}}
 \put(10,115){$T$}
 \put(195,5){$\mu$}
 \end{picture}
\caption{The deconfinement transition line in the $(\mu,T)$ plane. 
Phase transitions lines  dividing the plane in two regions, a hadron phase near the origin, and a deconfined phase beyond the curve. The red line(points) corresponds to the isotropic case, the blue line (points)  corresponds to the $E_x $ in the anisotropic case with $\nu=4$, the green to $E_y$ in the same  anisotropic case. The dots represent numerical data points, while the
solid interpolating lines are included "by hands". Red, blue and green lines correspond to $c=0.42$, darker red, blue and green lines correspond to $c=1.4$
}\label{fig:phasediag-y}
\end{figure}

 In Fig.\ref{fig:phasediag-y} we present the phase diagram for isotropic and anisotropic cases  for longitudinal 
 and transversal orientations of the quark pair.
  Holographic isotropic QCD phase diagram  has been studied previously in bottom-up approaches \cite{1008.3116,1201.0820,1506.05930}.
  We see that for zero chemical potential the deconfinement occurs for the low temperature in the anisotropic case, 
  and for near zero temperature the deconfinement  occurs for the larger chemical potential in the anisotropic casel.
  
 Fig.\ref{fig:phasediag-y} shows the dependence of the transition line on the orientation. Since quarks can be arbitraly oriented in respect to the collision line, this dependence on the orientation leads to a broadening  of the line separating  confinement  and deconfinement phases in the $(\mu,T)$-plane. It would be interesting to study this diagram for the anisotropic lattice.
 Decreasing of anisotropy decreases the broadening of the phase transition boundary. 

Let us remind that in isotropic case the confinement/deconfinment diagram has been studied in lattice QCD. Experimentally  the phase boundary between hadronic matter and the quark - gluon plasma in relativistic heavy ion collisions is probing using the HRM \cite{BraunMunzinger:1996mq,Andronic:2016nof}

\section{Direct photons and electric conductivity}
The thermal-photon production from the QGP plays an essential role, since photons after they are produced in HIC  almost do not interact with the QGP
and, therefore,  they give us  the local information in heavy ion collisions. 
The photon-emission rate is related to the retarded correlator of currents in momentum space
\cite{0607237}  
\begin{eqnarray}
G_{\mu\nu}^{R}(k)=i\int d^4(x-y)e^{ik\cdot(x-y)} \theta(x^0) \langle [ J^a_{\mu}(x), J^b_{\nu}(0) ] \rangle,
\end{eqnarray}
so that in the thermal equilibrium it is given by the light-like retarded correlator as
\begin{eqnarray}
d\Gamma=-\frac{d^3k}{(2\pi)^3}\frac{e^2n_b(|{\bf k}|)}{|{\bf k}|}\text{Im}\left[\text{tr}\left(\eta^{\mu\nu}G_{\mu\nu}^{ab \, R}\right)\right]_{k^0=|{\bf k}|},
\end{eqnarray}
where $\Gamma$ denotes the number of photons emitted per unit time per unit volume and $n_b(|{\bf k}|)$ denotes the thermal distribution function for bosons. 

The imaginary part of the retarded correlator is also related with the spectral function
\be
\chi _{\mu\nu}(k)=-2\Im\,G_{\mu\nu}^{R}(k)\ee
and due to the Kubo formula with  the conductivity tensor 
\be
\sigma _{\mu\nu}(k)=-\frac{G_{\mu\nu}^{R}(k)}{iw}\ee

The spectral function can be evaluated by holography using the flow method related with  the membrane paradigm \cite{Iqbal:2008by} . Below we shortly remind this prescription, see also \cite{1510.03321,1609.07208}.  One starts from the Maxwell action in the given background
\be
S_M=-{\cal N} \int d^5x \sqrt{-g}{V_{dil}(\phi )\over 4} F^{MN}F_{MN} \, ,
\ee
where  it is assumed also the presence of the  dilaton field $\phi$, and denote by ${\cal N}$ the normalization constant 
including the 5-dimensional gravitational constant $G_5$ etc.

The gauge field  $A_\mu (t,{\bf x},r)$  it is taken to be
\be
A_\mu (t,{\bf x},z) =\int {d^4 k \over (2 \pi)^4} e^{-i \omega t+i {\bf k x}}{\cal A}_\mu (z,\omega,k),\,\,\,\,\, {\cal A}_\mu (z,\omega,k)=
A_\mu (z,\omega,k)\,a_\mu (\omega,k)
\label{A-mu}
\ee
$A_\mu (t,{\bf x},z)$  satisfies the Maxwell equation, and $A_\mu (r,\omega,k)$ satisfies the boundary conditions at the boundary
\be
\lim _{z\to 0}A_\mu (z,\omega,k)=1\ee
and the infalling  boundary conditions at the horizon,
\be
\lim _{z\to z_h}A_\mu (z,\omega,k)=0\ee
One can consider for simplicity the case $k=k_x$, $k_{y_1}=k_{y_2}=0$. 

Let us parametrize the metric as 
\be\label{ds-aniz-con-chem-m}
ds^{2} = \frac{b^{2}(z)}{z^{2}}\left(-f( z)dt^{2} + dx^{2} + p(z)(dy^{2}_{1} + dy^{2}_{2})  + \frac{d z^{2}}{ f(z)}\right),
\ee
In our model $p(z)= z^{2-2/\nu}$. 
The Maxwell action (with the dilaton) has the form 
 \be\label{Max-ac} 
 \int d^4xdzV(z)\left(-\frac{F_{01}^2+F_{02}^2}{f 
   p }-\frac{F_{03}^2}{f }+f 
   \left(\frac{F_{1r}^2+F_{2r}^2}{p }+
  F_{3r}^2\right)-F_{0r}^2+\frac{F_{1
   2}^2}{p ^2}+\frac{F_{13}^2+F_{23}^2}
   {p }\right) \ee
 here
 \be
 V(z)=V_{dil}(\phi(z))\frac{b(z)}{z}\ee
 and 
 \be
 F_{MN}=\partial _MA_N-\partial _NA_M\ee
Assume that the fluctuating vector field depends only on $x_0,x_3,r$-coordinates, one can write the E.O.M.
 in term of $E_L$ and $E_\bot$
 
  \bea
   \label{EL}
E_L''+\left( \frac{ f'}{f}\frac{w^2}{w^2-fk^2}+   \frac{V'}{V}\right)\,E_L^\prime+ \frac{w^2-fk^2 }{f^2}\,E_L&=&0\\
 \label{EP}
    E_{\perp,i}'' + \left(
   -\frac{p' }{p}+
   \frac{V'}{V} +       \frac{f'}{f}\right) E_{\perp,i}'  +\frac{w^2  -k^2}{f}\,
    E_{\perp,i} & =&0,\,\,\,\,i=1,2
\eea
 here $
E_L=kA_0+wA_3$ and $E_{\perp,i}=w A_{\perp,i}$, $i=1,2$.

The boundary terms come from that terms the total action \eqref{Max-ac}
 which
have the z-derivative and we get 
\be\label{Sbound-sum}
S_{boundary}=S_{boundary,1}+S_{boundary,2}
\ee
where   $S_{boundary,1}$  and $S_{boundary,2}$  are
 \bea
 S_{boundary,1}&=&
\int d^4x
\,\frac{ f }{w^2-k^2f}\,E_LE_L^\prime
\\
S_{boundary,2} &=&\int d^4xV\,
\frac{f}{p\,w^2}E_{\perp}E'_{\perp},\,\,\,\,\,E_{\perp}E'_{\perp}\equiv\sum _{i=1,2} E_{\perp,i}E'_{\perp,i}
 \eea
In the analogy with the isotropic case we introduce
\bea
\label{zetaP-anis}
\zeta_\bot=-\frac{V \, f} {pw} { \partial_r E^\bot_i \over E^\bot_i} , \\
\label{zetaL-anis}\zeta_L=-{V \, f\,\over   \omega} { \partial_r E_L \over E_L},
\eea
which satisfy the following equations
\bea
 \label{zetaP-prime-aniso-f}
 \zeta_\bot'-\frac{w}{f}\left[\frac{p}{V}\,\zeta_\bot^2
+\frac{V}{p}\left(  1-f\frac{k^2}{w^2}\right)\right]&=&0.
\\
 \label{zetaL-prime-aniso-f}
\zeta'_L+\frac{k^2f'}{w^2-fk^2} 
 \zeta_L-\frac{w}{f}\left[\frac{\zeta_L^2}{V}+V\left(1-f\frac{k^2 }{w^2}\right)\right]&=&0
\eea

From these eqs follows that if $ \zeta_\bot'\neq \infty$ and $\zeta'_L \neq \infty$ then on the horizon should be
\bea
\left.\frac{p}{V}\zeta_\bot^2+\frac{V}{p}\right|_{z=z_h}&=&0\,\,\,\,\,\Rightarrow\,\,\,\,\,\zeta_\bot=i\frac{V(z_h)}{p}\\
\left.\frac{1}{V}\zeta_L^2+V\right|_{z=z_h}&=&0\,\,\,\,\Rightarrow\,\,\,\,\,\zeta_L=iV(z_h)
\eea

Since for $w=0$ the derivative  $\zeta_{\bot}'=0$ and we have 
\bea
\zeta_\bot|_{z=0}=\zeta_\bot|_{z=z_h}=i\frac{V}{p}|_{z=z_h}\eea

\begin{figure}[h!]
\centering
\begin{picture}(185,140)
\put(-100,0){\includegraphics[width=5cm]{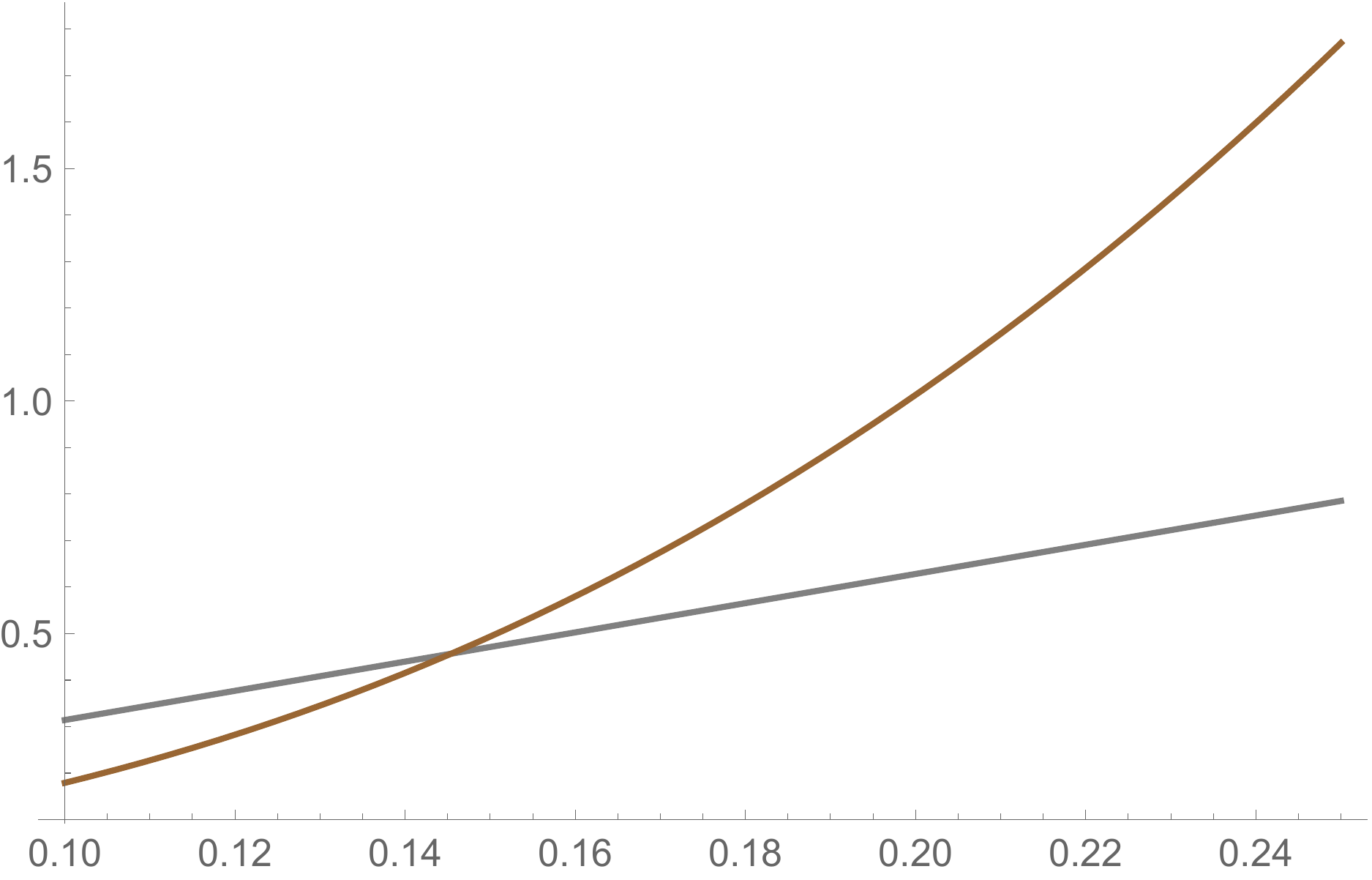}$\,\,\,\,\,\,a)$}
 \put(-100,100){$\sigma$}
 \put(40,5){$T$}
\put(120,0){\includegraphics[width=5cm]{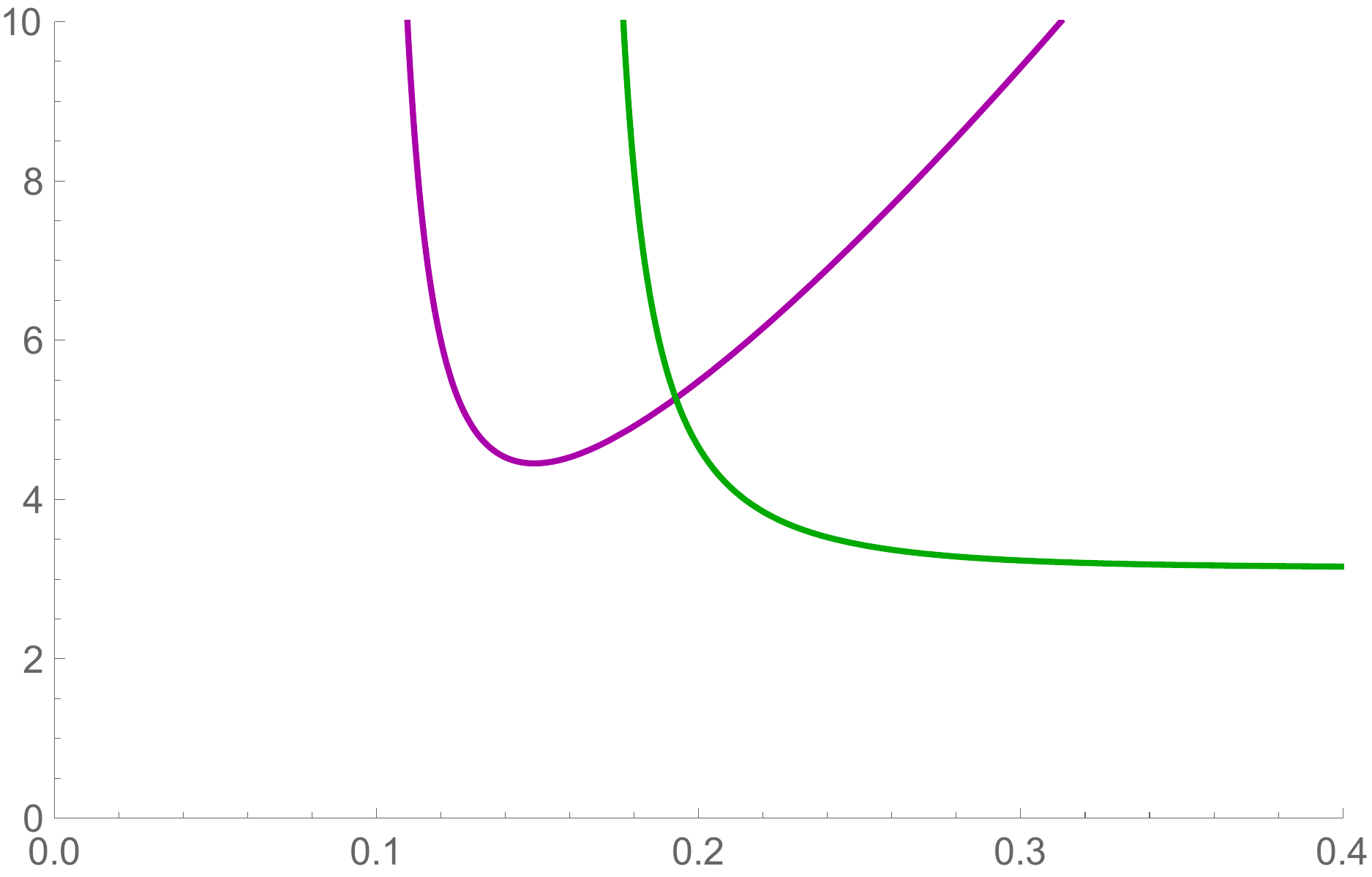}$\,\,\,\,\,\,b)$}
 \put(120,100){$\sigma$}
  \put(265,5){$T$}
\end{picture}
\caption{ The electric conductivity for a) $q=0$,  $\nu=1$ (the gray line) and $\nu=4$ (the brown line) 
and b) $q=0.2$,  $\nu=1$ (the magenta line) and $\nu=4$ (the green line)
} 
\label{fig:sigma-q}
\end{figure}
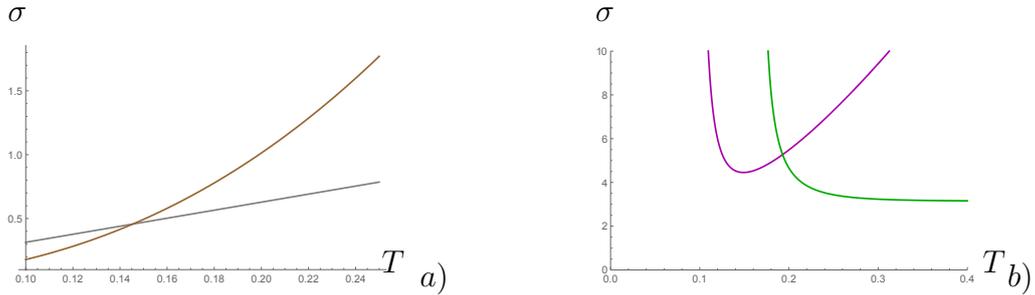

 $1/z_h$ and it dependence on the temperature is read from \eqref{temp} and approximately
 \be\label{temp-is-m}
\frac{1}{z_h}\approx
\frac{2\pi \nu}{1 + \nu}
\frac{T}{1 -(\frac{\nu+1}{\nu} )^{\frac{ 2+ 3\nu}{\nu}}q^2
 \left(
 \frac{1 }{2\pi T}\right)^{\frac{ 2+ 4\nu}{\nu}} }
\ee
In what follows we use this approximation. 
Therefore, the dependence of the electric conductivity on the temperature and the chemical potential  is given by 
\be
\label{zeta-nu}
|\zeta_\bot(T,\nu,q)|\approx\Big(\frac{2\pi \nu}{1 + \nu}\Big)^{3-2/\nu}
\frac{T^{3-2/\nu}}{\Big(1 -(\frac{\nu+1}{\nu} )^{\frac{ 2+ 3\nu}{\nu}}q^2
 \left(
 \frac{1 }{2\pi T}\right)^{\frac{ 2+ 4\nu}{\nu}}\Big)^{3-2/\nu} }
\ee
Form this formula we see, that increasing the anisotropy we increase the electric conductivity at hight temperatures, see 
 Fig.\ref{fig:sigma-q}.   We also  see that there is a critical temperature $T_0=T_0(\nu,q)$ such that for $T<T_0$ the conductivity decreases as we increase $\nu$. 

\begin{figure}[h!]
\centering \begin{picture}(185,150)
\put(0,0){\includegraphics[width=7cm]{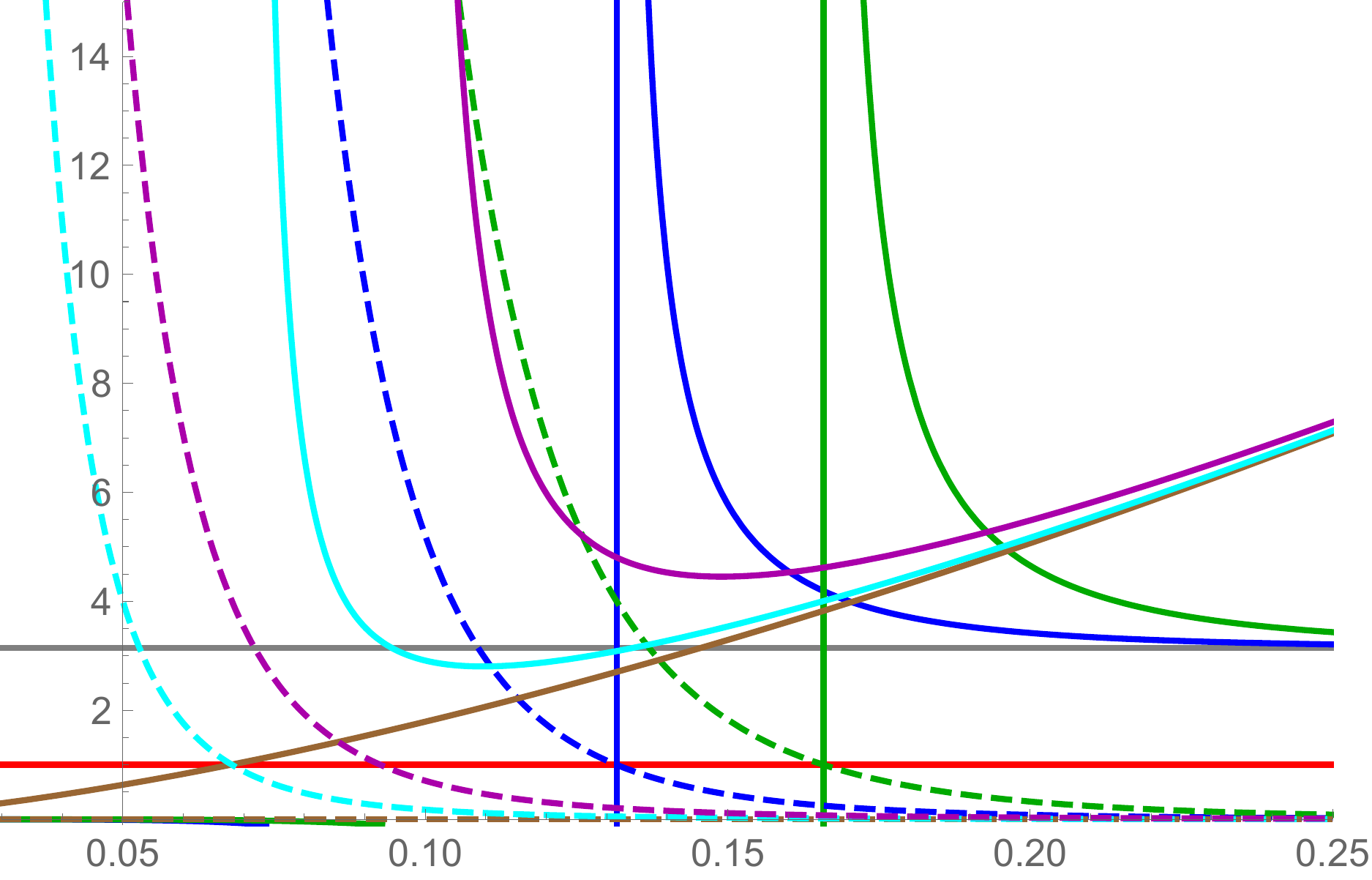}$\,\,\,\,\,\,$}
 \put(0,130){$\sigma/T$}
 \put(200,5){$T$}
\end{picture}
\caption{ The electric conductivity for  $\nu=1$ and $\nu=4$ and different values of $q$. The cases $\nu=1$ and $q=0,0.1,0.2$  are shown by gray, blue and green lines and
the cases $\nu=4$ and $q=0,0.1,0.2$  are shown by brown, darker cyan and darker green.  Dashed lines show validity of the approximation: only in the region where the dashed lines are below the red line
we can use this approximation.  
 }
\label{fig:sigma-T}
\end{figure}

\begin{figure}[h!]\begin{picture}(185,150)
\put(100,0){\includegraphics[width=7cm]{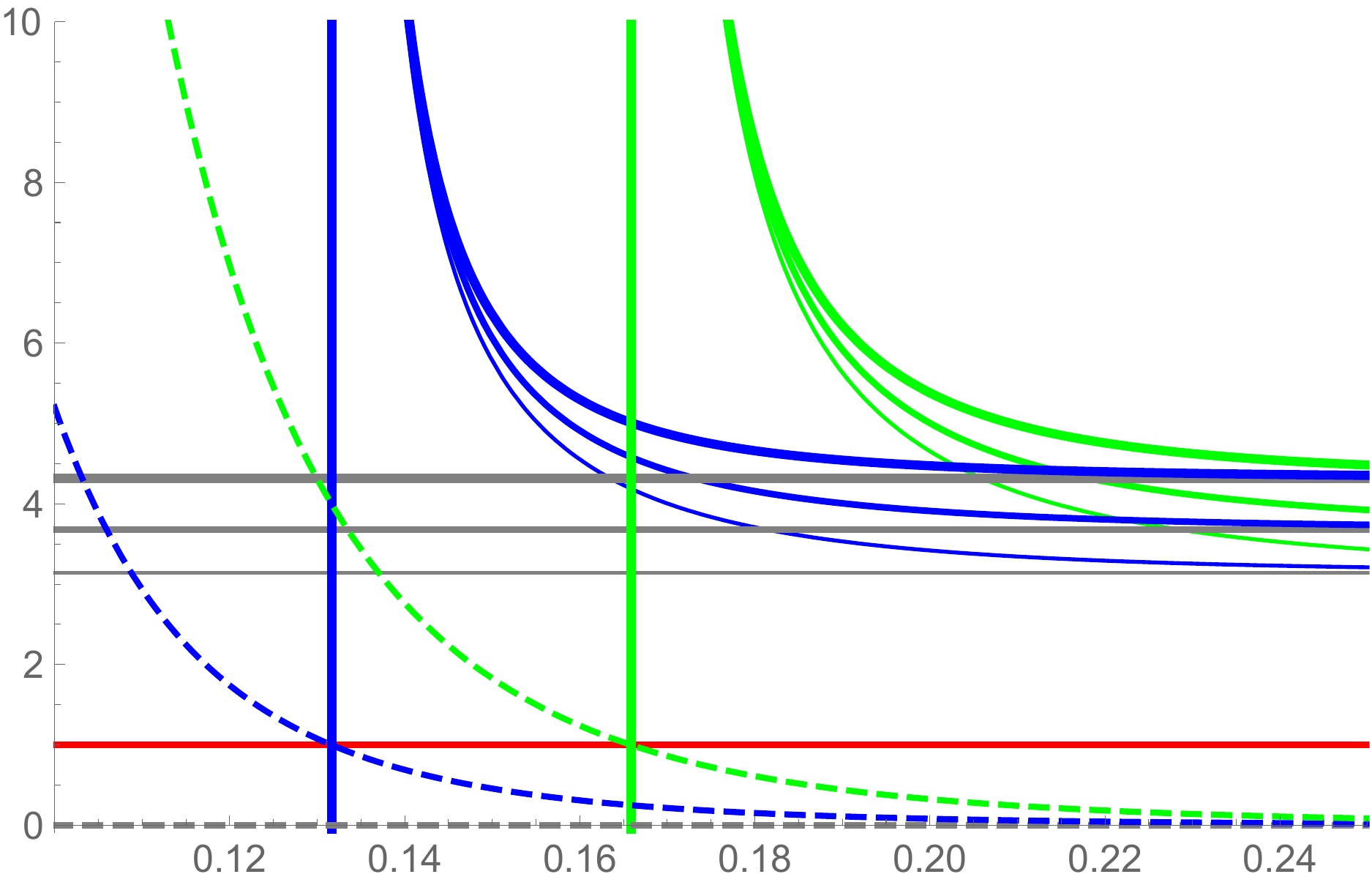}$\,\,\,\,\,\,$}
 \put(100,135){$\sigma/T$}
 \put(310,5){$T$}
\end{picture}\\
\caption{ The electric conductivity for  $\nu=1$, different values of $q$ and $c$. The gray solid lines show the 
electric conductivity for  $c=1$ and different values of $q=0,0.1,0.2$
(thin, middle and thick lines). The blue and green solid lines show the 
electric conductivity for  $c=1$ and  $c=1.2$, respectively, and different values of $q=0,0.1,0.2$
(thin, middle and thick lines). The   dashed lines show validity of the approximation: only in the regions where the dashed lines are below the red line
we can use our approximation.   
 }
\label{fig:sigma-T-c}
\end{figure}

Here we have to note that equation \eqref{zeta-nu}
 is valued only for $T>T_{app}$, see Fig.\ref{fig:sigma-T}.
In Fig.\ref{fig:sigma-T} and  Fig.\ref{fig:sigma-T-c} we plot the electric conductivity
as function of $T$ for  $\nu=1$ and  $\nu=4$  for different values of $q$ and the constant $c$ specifying the factor $b$. We see that  the confining factor $b(z)$ does not change  the qualitative picture much.

\begin{figure}[h!]
\begin{picture}(185,150)
\put(120,0){\includegraphics[width=7cm]{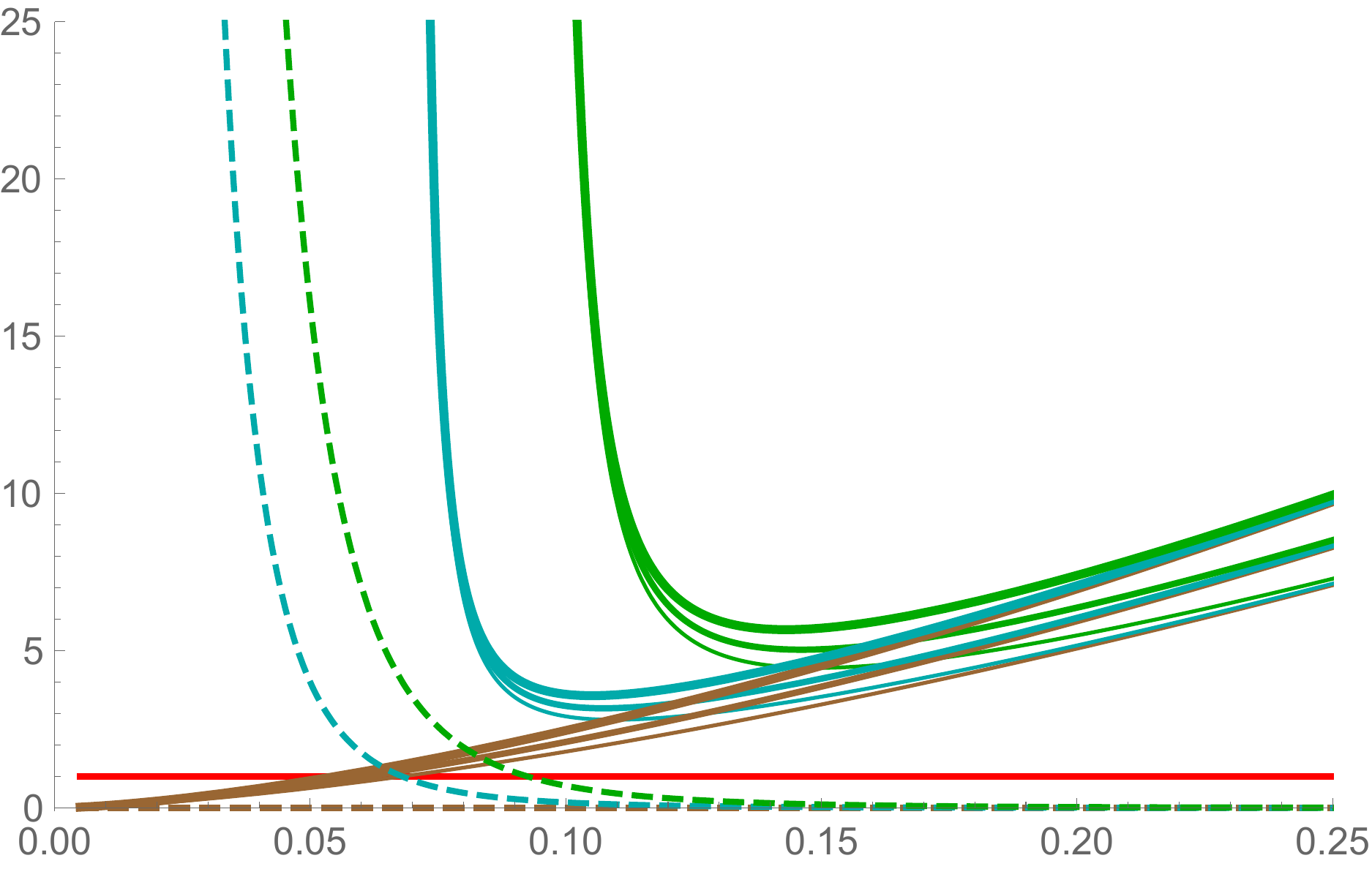}$\,\,\,\,\,\,$}
 \put(120,135){$\sigma/T$}
 \put(320,5){$T$}
\end{picture}
\caption{ The electric conductivity for  $\nu=4$ and different values of $q$ and  $c$. The brown solid lines show the 
electric conductivity for  $c=0$ and different values of $q=0,0.1,0.2$
(thin, middle and thick lines). The darker cyan  and green solid lines show the 
electric conductivity for   $c=1$ and  $c=1.2$ and different values of $q=0,0.1,0.2$
(thin, middle and thick lines).   The   dashed lines show validity of the approximation: only in the regions where the dashed lines are below the red line
we can use our approximation. 
 }
\label{fig:sigma-T-T-4}
\end{figure}

Note, that in the end of this section few comments are in order. Almost all 
 theoretical predictions  underestimate the direct-photon spectra. Several attempts \cite{McLerran:2015mda,Paquet:2015lta}, including the effects  from strong magnetic fields \cite{Muller:2013ila},   
have been undertaken
 to  fit  theoretical predictions to LHC experimental data. As it is  stressed in  \cite{1609.07208} direct photons calculations   from holography \cite{0607237,1210.7428,
 1305.5509,1609.07208} have to be complemented by  the medium 
 evolution and  the photons production from other phases. This  also concerns to our consideration.
 Here we have just presented a preliminary estimation of the role of the chemical potential, the Lifshitz type anisotropy and  the confining factor on the  photon emission rate. To give predictions for
  direct photons that can be observed at NICA one has to perform the study  similar to   \cite{1609.07208}. We have seen that at hight temperature the anisotropy of the Lifshitz type 
  increases the photon production and  the chemical  potential also increases it, meanwhile for temperature less than a critical one, they act in the opposite direction.

\section{Conclusion}

As a conclusion let me summarize. Motivated by the fact that 
to fit the experimental form of dependence of total multiplicity on energy we have to deal with a special anisotropic holographic model, related with the Lifshitz-like background,
we have estimated   the holographic  confinement/deconfinement  phase transition in the 
$(\mu,T)$  plane  in this anizotropic background. We have  found the dependence of the transition line on the orientation of the quark pair. This dependence leads to a non-sharp character of physical confinement/deconfinement phase in the $(\mu,T)$-plane. 
This calculation seems relevant  in the context of the future project NICA and FAIR.

We are also going to estimate  non-zero magnetic effects  in the the Lifshitz-like background,
in particular, by analogy with  \cite{Finazzo:2014zga}  one can estimate the crossover temperature in a magnetic field. 
As in isotropic case \cite{Rougemont:2015oea}  it is  interesting to find Debye screening mass near deconfinement in the anisotropic background.

\section*{Acknowledgments} I would like to thank the Organizers of ICNFP 2016 for
the kind invitation and warm hospitality.



\begin{thebibliography}{99}

 \bibitem{RHIC}
J. Adams  et al.  (STAR Collab.),
 Nucl.\ Phys.\  A {\bf 757}, 102 (2005)
\bibitem{Aamodt:2010pb}
   K. Aamodt   et al. (ALICE Collab.)
   Phys.\ Rev.\ Lett.  {\bf 105}, 252301 (2010)
   \bibitem{fluid2}
 E. Shuryak  
  Nucl.\ Phys.\  A {\bf 750}, 64 (2005)
\bibitem{Feinberg} E.L. Feinberg,   Phys. Usp.  {\bf 168}, 697  (1998)

\bibitem{Wilson} 
  K.~G.~Wilson at al,
  Phys.\ Rev.\ D {\bf 49}, 6720 (1994)

  \bibitem{Malda-rev}
O. Aharony  et al.
 Phys.\ Rept.  {\bf 323}, 183 (2000)

 \bibitem{PSS} G. Policastro, D. T. Son, and A. O. Starinets, 
 Phys.\ Rev.\ Lett.\  {\bf 87}, 081601 (2001)

\bibitem{solana}
 J.~Casalderrey-Solana, H.~Liu, D.~Mateos, K.~Rajagopal, U.~A.~Wiedemann,
  {\it Gauge/String Duality, Hot QCD and Heavy Ion Collisions}, (Cambridge University Press, 2014) 
  
\bibitem{NICA} V. Kekelidze et al, Nucl.Phys.A, {\bf 956},  846 (2016)

\bibitem{Bazavov:2015qsa} 
  A.~Bazavov,
  arXiv:1505.05543 [hep-lat].
\bibitem{Gursoy:2008za}
  U.~Gursoy, E.~Kiritsis, L.~Mazzanti and F.~Nitti,
  JHEP {\bf 05}, 033 (2009)

\bibitem{Gubser-rew} 
  O.~DeWolfe, S.~S.~Gubser and C.~Rosen,
  Phys.\ Rev.\ D {\bf 84}, 126014 (2011)


\bibitem{1108.6027}     G.~Aad et al. [ATLAS Collab.], 
  Phys.\ Lett.\ B {\bf 710}, 363 (2012)
  
  \bibitem{1512.06104} 
  J.~Adam {\it et al.} [ALICE Collab.],
  Phys.\ Rev.\ Lett.\  {\bf 116}, 222302 (2016)

\bibitem{IA}   
I.~Ya.~Aref'eva, 
  Phys. Usp.  {\bf 57},   527 (2014)

\bibitem{DeWolf}
O.~DeWolfe, S.~S.~Gubser, C.~Rosen and D.~Teaney,
  Prog. Part. Nucl. Phys. {\bf 75}, 86 (2014).

\bibitem{Wilke} 
  W.~van der Schee,
  arXiv:1407.1849 [hep-th]
\bibitem{Gubser} 
  S.~S.~Gubser, S.~S.~Pufu and A.~Yarom,
  Phys.\ Rev.\ D {\bf 78},  066014 (2008);
  JHEP {\bf 0911},  050 (2009)
\bibitem{AlvarezGaume:2008fx}
  L.~Alvarez-Gaume et al.,
  JHEP {\bf 0902},  009 (2009)

\bibitem{Lin:2009pn} 
  S.~Lin and E.~Shuryak,
  Phys.\ Rev.\ D {\bf 79},  124015 (2009)


\bibitem{Albacete:2009ji}
  J.~L.~Albacete, Y.~V.~Kovchegov and A.~Taliotis,
  JHEP {\bf 0905}, 060  (2009)
   \bibitem{ABG}
 I.~Y.~Aref'eva, A.~A.~Bagrov and E.~A.~Guseva,
 JHEP { \bf 0912}, 009 (2009)

\bibitem{ABJ}
 I.~Y.~Aref'eva, A.~A.~Bagrov and L.~V.~Joukovskaya,
 JHEP {\bf 1003} , 002, (2010)

\bibitem{Kovchegov:2009du}
  Y.~V.~Kovchegov and S.~Lin,
  JHEP {\bf 1003}, 057 (2010);
  Y.~V.~Kovchegov,
  Prog.\ Theor.\ Phys.\ Suppl.\  {\bf 187},  96 (2011)


\bibitem{KT} 
  E.~Kiritsis and A.~Taliotis,
  JHEP {\bf 1204}, 065 (2012)
  
  \bibitem{APP} 
  I.~Ya.~Aref'eva, E.~O.~Pozdeeva and T.~O.~Pozdeeva,
Teor.\ Mat.\ Fiz.\  {\bf 180}, 35 (2014).
  \bibitem{Ageev:2014mma} 
  D.~S.~Ageev and I.~Ya.~Aref'eva,
  J.\ Exp.\ Theor.\ Phys.\  {\bf 120}, 436 (2015)
  
 \bibitem{Landau}
 L.D. Landau,  Izvestya AN, Ser. Phys.{\bf 17}, 51 (1953)

\bibitem{Fermi} E. Fermi,  Progr. Theoret. Phys. {\bf 5}, 570 (1950)

\bibitem{Pomeranchuk}  I.Ya. Pomeranchuk, DAN SSSR {\bf 78}, 884 (1951) 


\bibitem{AG}
I. Ya. Aref'eva and A. A. Golubtsova, 
 JHEP {\bf 1504},  011 (2015)
 \bibitem{AGG}
 I.~Ya.~Aref'eva, A.~A.~Golubtsova and E.~Gourgoulhon,
  JHEP {\bf 1609}, 142 (2016)


  
\bibitem{9902170}  A. Chamblin, R. Emparan, C. V. Johnson and R. C. Myers, 
 Phys.\ Rev. D {\bf 60}, 064018 (1999)
\bibitem{1001.4414}  O.~Andreev,
  Phys.\ Rev.\ D {\bf 81} (2010) 087901
  
    \bibitem{1008.3116} 
 P.~Colangelo, F.~Giannuzzi and S.~Nicotri,
 Phys.\ Rev.\ D {\bf 83}, 035015 (2011);
 
   \bibitem{AZ}  
    O.~Andreev and V.~I.~Zakharov,
  Phys.\ Rev.\ D {\bf 74}, 025023 (2006)
\bibitem{ABP}   I.~Ya.~Arefeva, A.~A.~Bagrov and E.~O.~Pozdeeva,
  JHEP {\bf 1205}, 117 (2012)
  \bibitem{DiGi}
  D.~Giataganas,
  PoS Corfu {\bf 2012}, 122 (2013)
   \bibitem{Strickland:2013uga}  M. Strickland, 
 {\it Pramana}  {\bf 84},  671  (2015)
 
\bibitem{Aref'eva:2016doe} 
  I.~Aref'eva,
  EPJ Web Conf.,\  {\bf 125}, 01007 (2016)
\bibitem{IYaA}
I. Ya. Aref'eva, 
Theor. Math. Phys. {\bf 184},  1239 (2015)


 \bibitem{AAGG} D.~S.~Ageev, I.~Ya.~Aref'eva, A.~A.~Golubtsova and E.~Gourgoulhon,
  arXiv:1606.03995 [hep-th]

\bibitem{9803002}
J.~M.~Maldacena,
Phys.\ Rev.\ Lett.\  {\bf 80}, 4859 (1998)
\bibitem{9803135}  S.~J.~Rey, S.~Theisen and J.~T.~Yee,
  Nucl.\ Phys.\ B {\bf 527}, 171 (1998)
\bibitem{9803137}  A.~Brandhuber, N.~Itzhaki, J.~Sonnenschein and S.~Yankielowicz,
  Phys.\ Lett.\ B {\bf 434}, 36 (1998)
  \bibitem{Andreev:2006nw} 
O.~Andreev and V.~I.~Zakharov,
JHEP {\bf 0704}, 100 (2007)
\bibitem{Mia:2010zu}
  M.~Mia, K.~Dasgupta, C.~Gale and S.~Jeon,
Phys.\ Lett.\ B {\bf 694}, 460 (2011)

   
  
 \bibitem{1201.0820}   R.~G.~Cai, S.~He and D.~Li,
  JHEP {\bf 1203}, 033 (2012)
  
    \bibitem{1506.05930}  Y.~Yang and P.~H.~Yuan,
  JHEP {\bf 1512}, 161 (2015)
    \bibitem{Ewerz:2016zsx} 
  C.~Ewerz, O.~Kaczmarek and A.~Samberg,
  arXiv:1605.07181 [hep-th]


\bibitem{Fang:2015ytf} 
  Z.~Fang, S.~He and D.~Li,
 Nucl.\ Phys.\ B {\bf 907}, 187 (2016)  
 
  
\bibitem{BraunMunzinger:1996mq} 
  P.~Braun-Munzinger and J.~Stachel,
  Nucl.\ Phys.\ A {\bf 606}, 320 (1996)
  \bibitem{Andronic:2016nof} 
  A.~Andronic, P.~Braun-Munzinger, K.~Redlich and J.~Stachel,
  arXiv:1611.01347 [nucl-th].
  
   \bibitem{0607237} S.~Caron-Huot, P.~Kovtun, G.~D.~Moore, A.~Starinets and L.~G.~Yaffe,
  JHEP {\bf 0612}, 015 (2006)

   \bibitem{Iqbal:2008by} 
  N.~Iqbal and H.~Liu,
  Phys.\ Rev.\ D {\bf 79}, 025023 (2009)
  
\bibitem{1510.03321}   S.~I.~Finazzo and R.~Rougemont,
  Phys.\ Rev.\ D {\bf 93},  034017 (2016)
  \bibitem{1609.07208} 
  I.~Iatrakis, E.~Kiritsis, C.~Shen and D.~L.~Yang,
  arXiv:1609.07208 [hep-ph]

\bibitem{1611.04848}   I.~Iatrakis, E.~Kiritsis, C.~Shen and D.~L.~Yang,
  arXiv:1611.04848 [hep-ph]
  
   \bibitem{McLerran:2015mda} 
  L.~McLerran and B.~Schenke,
  Nucl.\ Phys.\ A {\bf 946}, 158 (2016)
  \bibitem{Paquet:2015lta} 
  J.~F.~Paquet et al,
  Phys.\ Rev.\ C {\bf 93}, 044906 (2016)
 
 \bibitem{Muller:2013ila} 
  B.~Muller, S.~Y.~Wu and D.~L.~Yang,
  Phys.\ Rev.\ D {\bf 89},  026013 (2014)

    \bibitem{1210.7428}   K.~A.~Mamo,
  JHEP {\bf 1308}, 083 (2013)
  
  \bibitem{1305.5509}   S.~Y.~Wu and D.~L.~Yang,
  JHEP {\bf 1308}, 032 (2013)
  
 \bibitem{Finazzo:2014zga} 
  S.~I.~Finazzo and J.~Noronha,
  Phys.\ Rev.\ D {\bf 90}, 115028 (2014)
  
\bibitem{Rougemont:2015oea} 
  R.~Rougemont, R.~Critelli and J.~Noronha,
  Phys.\ Rev.\ D {\bf 93}, 045013 (2016) 


\end{thebibliography}
\end{document}